\documentclass{iucrjournals}
\usepackage{amsmath}
\usepackage{comment}
\usepackage{subcaption}
\usepackage{hyperref}

\title{Approximate resolution convolution function for fitting a dispersion gap measured on a triple-axis spectrometer}

\author[a,b]{Emma Y. Lenander\IUCrCemaillink{emma.lenander@nbi.ku.dk}\IUCrOrcidlink{0000-0002-1274-3579}}%
\author[a]{Silas B. Schack\IUCrOrcidlink{0009-0001-8057-7859}}
\author[a]{Kim Lefmann\IUCrOrcidlink{0000-0003-4282-756X}}
\author[b,a]{Henrik M. Rønnow\IUCrOrcidlink{0000-0002-8832-8865}}%

\affil[a]{Niels Bohr Institute, University of Copenhagen, DK-2100 Copenhagen \O , Denmark}
\affil[b]{Institute  of  Physics,  \'Ecole  Polytechnique  F\'ed\'erale  de  Lausanne  (EPFL),  CH-1015  Lausanne, Switzerland}

\begin{document} 
\maketitle 

\begin{synopsis}
We present an analytic convoluted-gap function for fitting a both a linear or quadratic dispersion gap in a constant-Q cut measured with a focused triple-axis-spectrometer. It outperforms previous methods of fitting a gap, without doing a full resolution convolution.
\end{synopsis}

\begin{abstract}

We present an analytic convoluted-gap function, eq. \eqref{eq:analytical}, for fitting dispersion gaps measured on triple-axis spectrometers (TAS). At the gap, the instrumental resolution skews the signal, producing a high-energy tail that complicates fitting. Our function assumes an instrumental $Q$-resolution with two equal wide and one narrow direction (typical of focused TAS instruments), and a parabolic dispersion at the gap, which is exact for quadratic and accurate for linear dispersions if the resolution is moderate. We demonstrate, that our function outperforms previous methods of fitting a gap, by giving a better fit and more accurate gap determination, seen in figure \ref{fig:McStats_fits}. Here, the anti-ferromagnetically gapped material; MnF$_2$ is simulated in a double-focusing TAS instrument. 
We also tested our function on experimental data on MnF$_2$ from a TAS-like instrument, where we reproduce the gap size from previous accurate experimentally determined measurements.
The function is simple to implement, converges reliably, and we recommend its use for future gap fitting on TAS data.
\end{abstract}


\section{Introduction}

Neutron spectroscopy is a very powerful technique to investigate magnetic and lattice dynamics in materials as function of momentum and energy transfer. The neutron scattering cross-section is proportional to the spatial and temporal Fourier transform of the respective atomic displacement and spin-spin correlation functions\cite{boothroyd_2020}, embodied in the dynamical structure factor $S(\textbf{Q},\omega)$. In many cases, the ground state spontaneously breaks a continuous symmetry of a system, resulting in gapless symmetry-restoring excitations (Goldstone modes). In crystals, the lattice dynamic is well described by phonons, and in magnetically ordered systems the primary magnetic excitations are spin waves. Phonons restore the broken translational symmetry of a crystal, and spin-waves restore the rotational spin symmetry in Heisenberg and XY models. In both cases, these coherent collective excitations give rise to dispersive bands of momentum dependent excitation energies with corresponding structure factors: 
\begin{equation}
    S(\textbf{Q},\omega)=\sum_n S_n(\textbf{Q}) \delta(\omega-\omega_n(\textbf{Q})),
    \label{eq:S(Q,omega}
\end{equation}
where $\omega_n(\mathbf{Q})$ denotes the dispersion relation of the $n$th mode and $S_n(\mathbf{Q})$ its corresponding spectral weight\cite{Shirane}. The $\delta$-function enforces energy conservation, such that each branch contributes only at its characteristic excitation energy. 
The dispersion of the Goldstone modes close to zero energy can be linear (e.g. acoustic phonons and antiferromagnetic spin waves) or quadratic (e.g. ferromagnetic spin waves and triplon excitations). For a quadratic excitation, $S_n(\textbf{Q})$ is constant, while for a linear excitation, $S_n(\textbf{Q}) \propto 1/\omega$. 
\\
Various effects including anisotropy, pinning, incomplete softening or long-range interactions can gap these excitations. Detecting and quantifying such gaps reveal some of the key information that can be extracted from spectroscopic investigations.
A gapped Goldstone mode is like a normal Goldstone mode, only shifted up in energy. At vanishing momentum, $|\textbf{Q}-\textbf{Q}_0| \sim 0$, the excitation does not have zero energy, but the dispersion curve starts at a finite gap, $\Delta$. For a quadratic gapped dispersion, $E_{quadratic}(\textbf{Q})$, the Goldstone mode will follow
\begin{equation} 
E_{quadratic}(\textbf{Q}) = \alpha(\textbf{Q}-\textbf{Q}_0)^2+\Delta,
\label{eq:parabola}
\end{equation}
where $\alpha$ determines the curvature of the parabola. In the other scenario, with a linear dispersion becoming gapped, the dispersion is given by,  
\begin{equation} 
    E_{linear}(\textbf{Q}) = \sqrt{(a(\textbf{Q}-\textbf{Q}_0))^2+\Delta^2},
    \label{eq:AFM_dispersion}
\end{equation} 
where $a$ is the spin-wave velocity (the slope of the dispersion at \textbf{Q}-values away from the gap region). We here assume that the spin wave velocity is identical for all crystallographic directions. In the limit $\textbf{Q} \rightarrow\textbf{Q}_0$ ($a(\textbf{Q}-\textbf{Q}_0) \ll \Delta$), we perform a series expansion of eq. \eqref{eq:AFM_dispersion}, giving the parabolic form of eq. \eqref{eq:parabola} with $\alpha=a^2/(2\Delta)>0$.

\begin{figure}[ht!]
    \centering
    \includegraphics[width=0.6\linewidth]{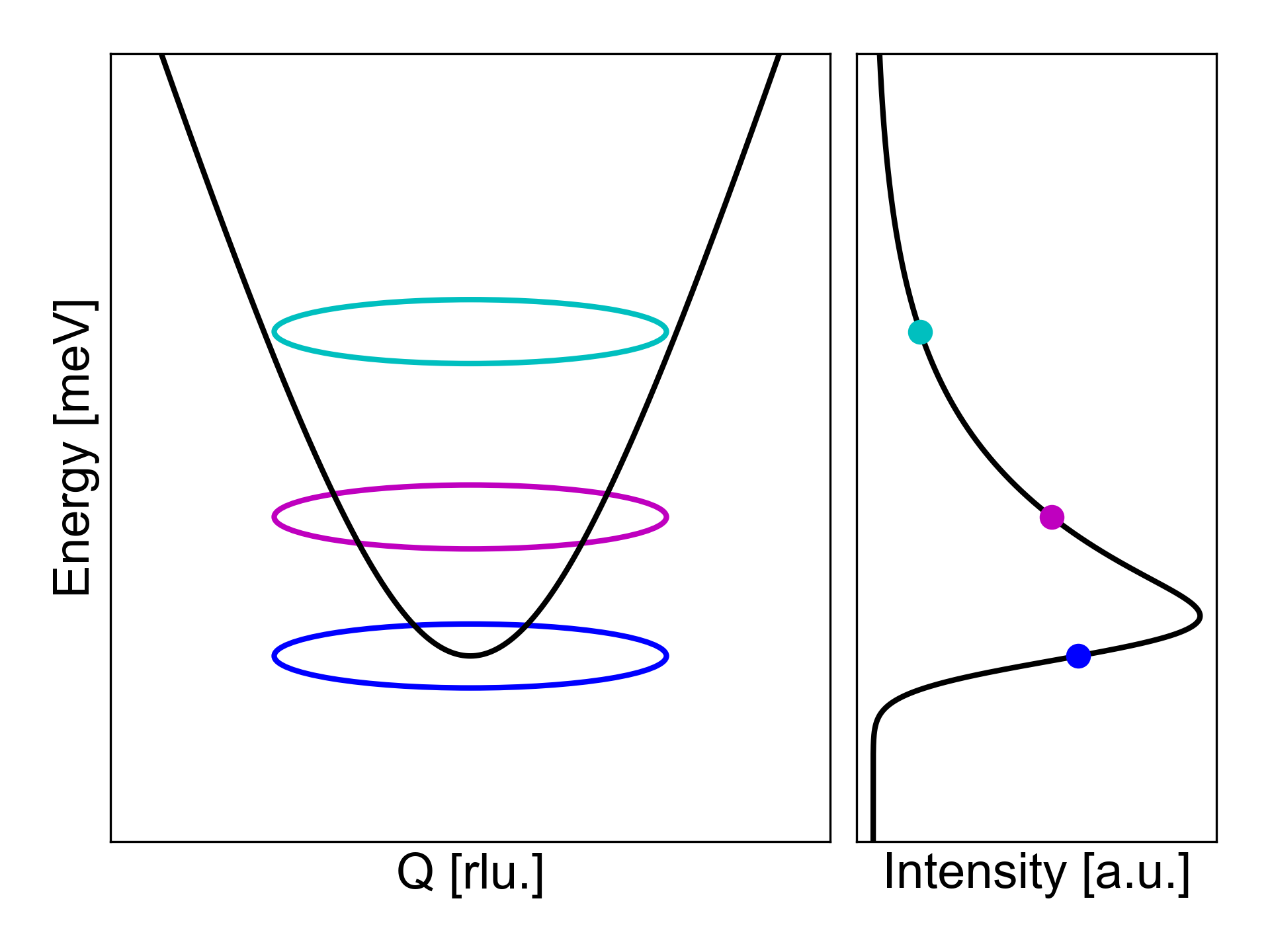}
    \caption{(Left) The linear dispersion in eq. \eqref{eq:AFM_dispersion}, energy transfer as a function of \textbf{Q} (black line). The instrumental resolution 
    at the gapped \textbf{Q}-position is drawn as coloured ellipsis at varying energy transfers. (Right) The intensity in a constant Q-cut at the gapped position, illustrating how the high-energy tail above the spin gap appears from the instrumental resolution. }
    \label{fig:resolution}
\end{figure}

In neutron spectroscopy, the measured signal is not an ideal $\delta$-function, but is broadened by the finite resolution of the instrument. This broadening is described by the resolution function $G(\mathbf{Q}-\mathbf{Q}', \omega-\omega')$, which represents the spectrometer’s response in both momentum and energy space. The respective $Q$- and $E$-resolutions determine how finely the momentum and energy dependence of the excitation can be resolved: limited $Q$-resolution broadens and smooths the apparent dispersion in reciprocal space, while limited $E$-resolution smears the energy profile, rounding spectral features and filling in sharp gaps. The observed intensity is thus given by the convolution of the intrinsic scattering function $S(\mathbf{Q}, \omega)$ with the instrumental resolution:
\begin{equation}
    I(\mathbf{Q}, \omega) \propto \int G(\mathbf{Q}-\mathbf{Q}', \omega-\omega')\, S(\mathbf{Q}', \omega') \, d\mathbf{Q}'\, d\omega'.
    \label{eq:intensity}
\end{equation}
This convolution broadens and, in some cases, skews the theoretical $\delta$-function peaks in $S(\mathbf{Q}, \omega)$, producing the experimentally observed intensity distribution, see figure \ref{fig:resolution}. To obtain the exact cross-section, one must also include the fundamental constants, the dipole factor, and the ratio $k_f/k_i$.

\begin{figure}[ht!]
    \centering
    \includegraphics[width=0.5\linewidth]{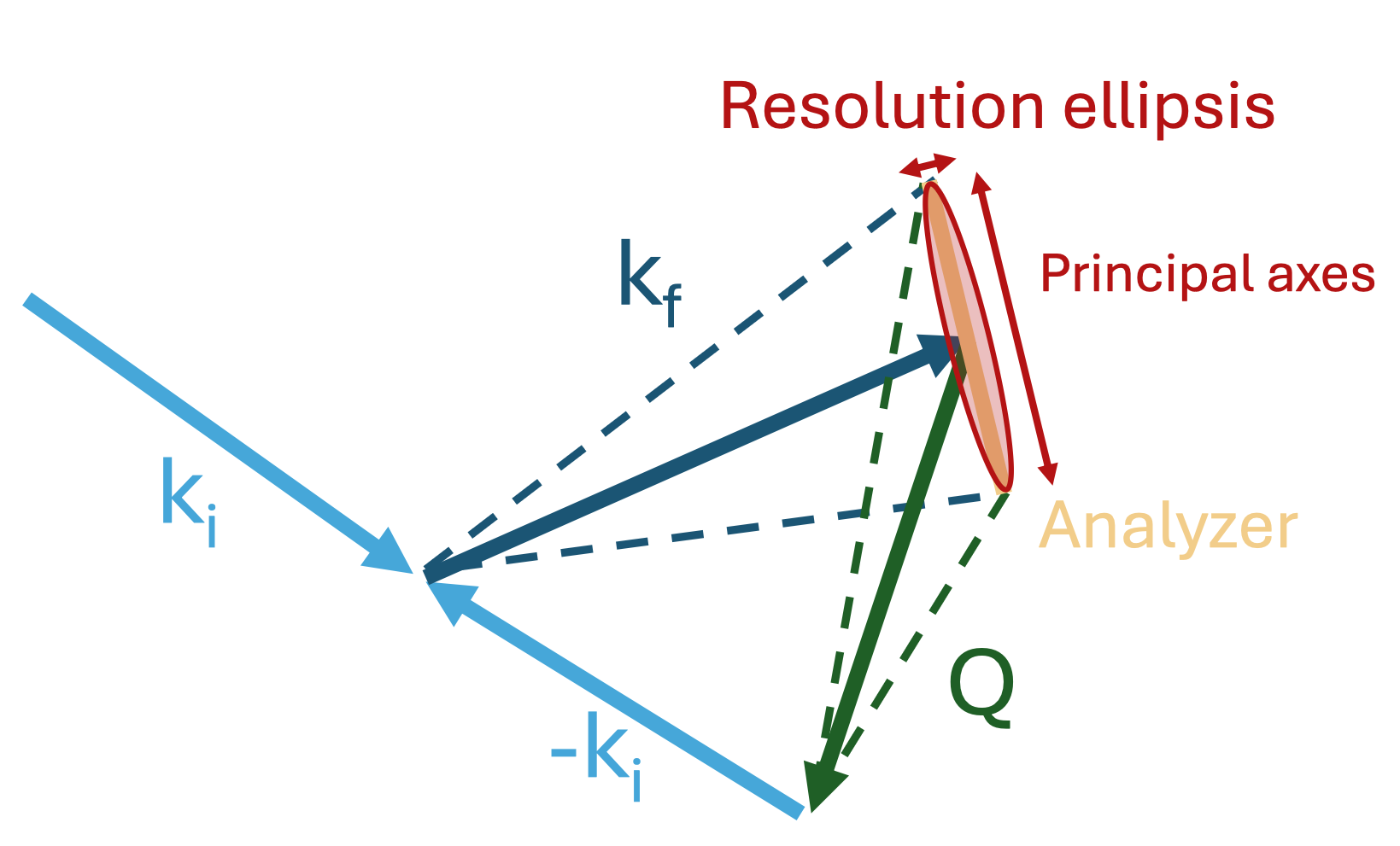}
    \caption{The 2-dimentional part of the $Q$-resolution ellipsoid for a triple-axis-spectrometer is indicated in red. The scattering vector $\textbf{Q}=\textbf{k}_f -\textbf{k}_i$, with the size of the analyzer, determines the 3D $Q$-resolution, which can be split into 3 components; two $dQ$ principal axis in the scattering plane; one narrow and one wide, and $dQ_{vert}$, which is the resolution out of the scattering plane (wide, and not drawn). The instrument also has an energy resolution, which needs to be taken into account.}
    \label{fig:Resolution_ellipsis}
\end{figure}

The shape and width of the resolution function depend on several factors, including the incident and final neutron wave vectors ($\textbf{k}_i$ and $\textbf{k}_f$), the mosaic spreads of the monochromator and analyzer crystals, the collimation setup, the distances between instrument components, the scattering angles, and the wavelength spread of the neutrons, see figure \ref{fig:Resolution_ellipsis}. Thus, a full resolution calculation requires detailed knowledge of the experimental configuration, 
and information about the sample properties may also be necessary to properly model the scattering volume. Therefore, the resolution function is typically approximated by a Gaussian in four-dimension; \textbf{Q} and $\hbar\omega$, that incorporates contributions from various components of the instrument \cite{Shirane}.

In practice, the form and anisotropy of the resolution function vary between spectrometer types. One typically measures dynamics on a time-of-flight (TOF) spectrometer or a Triple-Axis Spectrometer (TAS). A pixelated TOF spectrometer often has relatively fine $Q$-resolution, which can be tuned after the experiment by balancing bin-width versus statistics. 
In contrast, a focusing TAS generally has wider $Q$-resolution in two of three directions, one of these being the vertical direction. For TAS, the instrument resolution can be determined analytically using the Cooper–Nathans \cite{Cooper_1967} and Popovici \cite{Popovici_1975} formalisms to compute the resolution matrix $\mathbf{M}$ under the assumption that all uncertainties have Gaussian distributions and linear approximations around the nominal instrument settings. In this framework, the resolution function can be written as
\begin{equation}
    G(\mathbf{Q}, \omega) \sim \exp\!\left[-\tfrac{1}{2}\, \Delta\mathbf{X}^\mathrm{T}\, \mathbf{M}^{-1}\, \Delta\mathbf{X}\right],
\end{equation}
where $\Delta\mathbf{X} = (\Delta\mathbf{Q}, \Delta\omega)$. The resolution matrix can be computed numerically by Monte Carlo integration, as implemented in software packages such as \textsc{Takin} \cite{Takin2023}, or through ray-tracing simulations using \textsc{McStas} \cite{McStas_1999, Lefmann_2000, McStas_2020}. One can also determine it experimentally by scanning single-crystal or powder Bragg peaks, 
see a guide in appendix \ref{app:scanning_Bragg_peak}.

\begin{figure}[ht!]
     \centering
         \centering
         \includegraphics[width=0.4\textwidth]{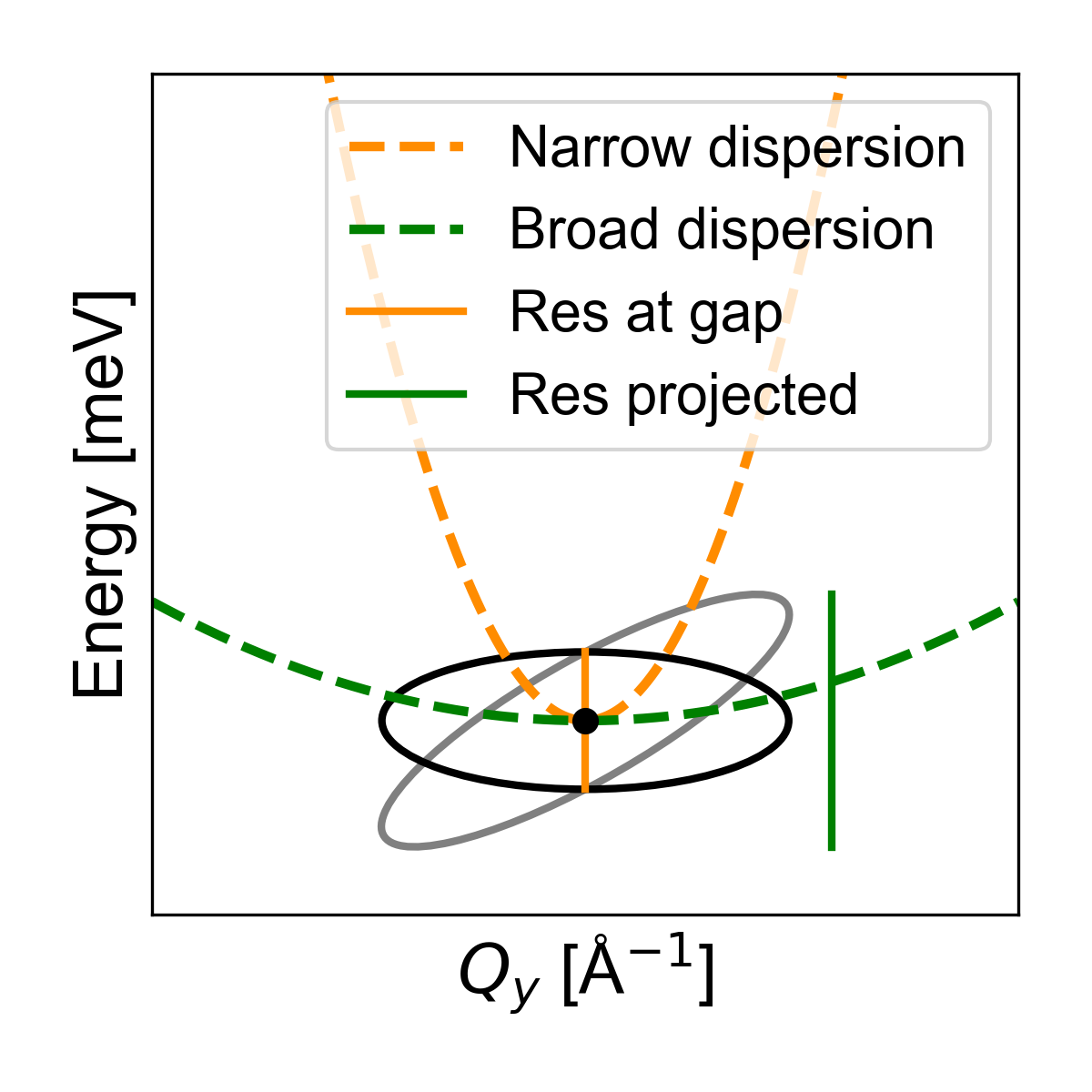}
        \caption{The intrinsic instrumental energy resolution at a specific \textbf{Q} is shown in orange and the projected effective resolution in green. For a broad dispersion, one should use the projected resolution (green), while for a narrow dispersion, the energy resolution at a specific \textbf{Q} is more correct. For our approximation, we do not see the correlations between Q and E as in the real case (grey ellipsis), so we assume an ellipsis without a tilt and with the energy resolution depending on the steepness of the dispersion. The black ellipsis indicates the case of a narrow dispersion (orange).}
        \label{fig:resolution_assumptions}
\end{figure}

In this manuscript, we consider the case of an instrument with a broad $Q$-resolution in two of the three reciprocal-space directions, representative of a focused TAS setup. The remaining direction, with the shortest principal axis of the ellipsoid, is approximated by a delta function, $\delta(Q_x)$. Here, $x$ is in general not aligned with the crystallographic axis, but this is unproblematic in our isotropic approximation. The other two principal axes are modelled by Gaussian resolution functions of equal width, denoted $\sigma_Q$. The energy resolution is represented by a Gaussian of width $\sigma_E$. The assumptions of the resolutions are the $Q$-resolution has the shape of a pancake and the energy resolution depends on the steepness of the dispersion. For a narrow dispersion compared to the resolution ellipsis, indicated in orange in figure \ref{fig:resolution_assumptions}, one will use the intrinsic instrumental energy resolution at a specific \textbf{Q} (at the gap). For a broad dispersion, one should use the effective projected energy resolution (green). Under these assumptions, the total resolution function can be approximated as
\begin{equation}
    G(\textbf{Q},\omega) \approx \delta(Q_x) \exp \left( -\frac{
    Q_y^2+Q_z^2}{2 \sigma_Q^2} - \frac{\hbar^2\omega^2}{2{\sigma_E^2}}\right).
    \label{eq:resolution_approx}
\end{equation}
This is one of the approximations, that we will make in order to derive an analytical function for the resolution convoluted scattering function.

To experimentally measure the gap of the dispersion with a TAS, one usually performs a constant-Q scan (or cut) at the bottom of the dispersion. Thus, the data consist of intensity ($I(\textbf{Q},\omega)$) as a function of energy transfer ($\hbar \omega$). Taking the instrument resolution into account, the approximation that the dispersion follows a parabola at the gap, eq. \eqref{eq:AFM_dispersion}, is valid for the linear excitation when the series expansion is valid or when the $Q$-resolution is narrower than the sides of the parabola. If the energy variation of the linear dispersion is within the $Q$-resolution range and the instrument’s energy resolution are both small compared to the energy gap, then $1/\omega_Q$ can be approximated as constant, like the quadratic excitation. A gapped dispersion can, therefore, be described by eq.~\eqref{eq:parabola} and $S_n(\textbf{Q})$ is constant. This results in the scattering function in eq.~\eqref{eq:S(Q,omega} being written as: 
\begin{equation}
    S(\textbf{Q},\omega) \propto \delta(\hbar\omega-(\alpha(\textbf{Q}-\textbf{Q}_0)^2 + \Delta) ).
    \label{eq:delta_func_approx}
\end{equation}
This is our second approximation, which will be used in order to derive the analytical function for the resolution convoluted scattering function.

Normally to determine the gap size of a constant-Q scan, one will typically fit the dispersion with a Gaussian line-shape and a constant or linear background. This works well away from the gap, but at the bottom of the dispersion, due to the instrument resolution, the scan tends to pick up signal from higher energies. This results in a tail at energies above the gap, see figure \ref{fig:resolution}, which is not captured in a Gaussian line-shape. To get a more precise value for the gap position, one can use a function that captures the tail. A typical example of such a function, is a Gaussian line-shape for the low-energy part of the peak and a Voigt for the high-energy part of the peak (see appendix \ref{app:Gauss+Voigt}). The Voigt profile is a convolution of a Gaussian and a Lorentzian function, thus it has a Gaussian peak shape with Lorentzian tail. This function has the potential to fit better, but it is not a physical description of the resolution, and the function tends to overestimate the gap size. 

In this manuscript, we propose an approximate function to fit the cut at the gap to a skewed function: a convoluted-gap model (see eq. \eqref{eq:analytical}). Here we use the two approximations mentioned above; that the gapped dispersion follows a parabola at the gap position and that the $Q$-resolution has two equally wide directions and one vanishing narrow, typical for a focusing TAS.
The function also includes the energy resolution and the spin wave velocity of the gapped dispersion. By simulating an AFM gapped material, MnF$_2$, in a double-focusing TAS instrument, we show that our function performs much better than the Gaussian line-shape and the combined Gaussian-Voigt function. We have also tested our function on experimental data on MnF$_2$ from the TAS-like instrument CAMEA (PSI). Even in the case of a TAS-like instrument, which has a more complicated resolution, we find that our function fits the data well and succeeds in finding the gap value.

\section{Analytical resolution convoluted function for fitting a gap}\label{sec:fiiting_func}

Using the approximations presented in the previous section, eqs.~\eqref{eq:resolution_approx} and \eqref{eq:delta_func_approx}, we can write the intensity given by eq.~\eqref{eq:intensity} as:
\begin{eqnarray}
I(\mathbf{Q}, \omega)  &\propto & \int_{-\infty}^\infty \delta(Q_{x0} - Q_x') \exp \left( -\frac{ (Q_{y0} - Q_y')^2 + (Q_{0z} - Q_z')^2}{2 \sigma_Q^2} - \frac{\hbar^2(\omega - \omega')^2}{2 \sigma_E^2} \right) \nonumber \\
 & \times & \delta(\hbar\omega' - \Delta - \alpha (\textbf{Q}-\textbf{Q}_0)^2) \; d\mathbf{Q'} \, d\omega'.
\end{eqnarray}
From this expression for the intensity, we present an analytical solution, the details of the derivation being described in appendix \ref{app:derivation_numerical}. We find that the intensity is described as a convolution of the energy resolution with an asymmetric function $f(\omega)$ that includes the $q$-part of the resolution. Since $f(\omega)$ only picks up intensity above the gap, $\hbar\omega>\Delta$, the function includes the Heaviside function, $H$:
\begin{equation}
    f(\omega) = H(\hbar \omega - \Delta) \exp \left( -\frac{|\hbar\omega - \Delta|}{2\alpha \sigma_Q^2} \right) =
\begin{cases}
\exp \left( -\frac{|\hbar\omega - \Delta| }{2 \alpha \sigma_Q^2} \right), & \hbar\omega > \Delta . \\
0, & \hbar\omega \leq \Delta .
\end{cases}
\label{eq:heavyside}
\end{equation}
From this, the final intensity becomes:
\begin{equation}
    I(\mathbf{Q}, \omega) = A 
    \frac{1}{\sqrt{2 \pi \sigma_E^2}} \frac{1}{2 \alpha \sigma_Q^2}
    \int_{-\infty}^\infty f(\omega') \exp \left( - \frac{(\hbar\omega - \hbar\omega')^2}{2 \sigma_E^2} \right) \, d\omega' + B.
    \label{eq:fitting_func}
\end{equation}
The Heaviside function in $f(\omega)$ restricts the domain of integration to 
$\hbar \omega \geq \Delta$, effectively changing the integration interval to $[\Delta,\infty)$.
The Gaussian normalisation constants are written explicitly, while all other experimental constants e.g. the prefactor of $S(\textbf{Q}, \omega)$ and the intensity normalisation of the instrument are described by a normalisation constant, $A$. To account for the instrument background, the scalar, $B$, is used. 

Eq. \eqref{eq:fitting_func} is to be determined numerically, which makes it in practice more difficult for a fitting routine to converge at the global minimum. Thus, we are interested in finding an analytical expression (see appendix \ref{app:derivation_analytical}). The Heaviside function in $f(\omega)$ in eq.~\eqref{eq:heavyside}, ensures support only for $\hbar\omega \ge \Delta$, but the convolution itself yields a smooth non-zero tail for all $\hbar\omega$ given by the erfc-function. The analytical convoluted-gap function is finally written by:
\begin{equation}
I(\textbf{Q}, \hbar\omega)
= A
\frac{1}{4 \alpha \sigma_Q^2}\;
\exp\!\Big[\frac{1}{2\alpha \sigma_Q^2}(\Delta - \hbar\omega) + \tfrac{1}{2} \Big(\frac{ \sigma_E}{2 \alpha \sigma_Q^2} \Big)^2\Big]\;
\operatorname{erfc}\!\left(\frac{\Delta - \hbar\omega + \Big(\frac{ \sigma_E}{2 \alpha \sigma_Q^2} \Big)^2}{\sqrt{2}\,\sigma_E}\right) +B
\label{eq:analytical}
\end{equation}

If the approximation of the $Q$-resolution holds, the analytical function in eq. \eqref{eq:analytical} is exact for gapped quadratic dispersions, like FM spin waves. It is valid for gapped linear excitations, like phonons or an AFM spin waves, under the assumptions that the excitation at the gap follows a parabola and that the factor $S_n(\textbf{Q})\propto 1/\omega$ is constant. For the linear excitation, one can replace $\alpha=a^2/(2\Delta)$. For more details on the assumptions, see appendix \ref{app:limitations}.

\section{Simulated data; \texorpdfstring{MnF$_2$}{MnF2}}
To demonstrate the use of our expression for the analytical resolution convoluted-gap function, eq.~\eqref{eq:analytical}, we choose to study examples from the antiferromagnet MnF$_2$. This material has been extensively studied as a model system in magnetism, particularly through neutron scattering. Its simple structure and well-defined magnon dispersion makes it a benchmark material for understanding AFM spin dynamics \cite{Yamani_2010}. MnF$_2$ is a well-known AFM insulator that crystallizes in the tetragonal rutile structure (space group \textit{P4$_2$/mnm}). It has a Néel temperature of approximately 67~K, below which Mn$^{2+}$ ions ($S=5/2$) align antiparallel along the $c$-axis \cite{Erickson_1953}, forming a collinear antiferromagnetic order. Its magnon dispersion is gapped, with a spin-wave gap of approximately 1.08 meV at the zone center \cite{Johnson_1959, Low_1964, Nikotin_1969}. Thus, this material is a good model example for this work. Recently, MnF$_2$ has attracted renewed interest, because it has been classified as an altermagnet \cite{Smejkal2022}, supposedly exhibiting spin-dependent band structure symmetries without net magnetization. Previous experimental studies for the altermagnetic splitting of the magnon modes in MnF$_2$ were unsuccessfully \cite{Morano_2025}, however, a recent study revealed the splitting using polarised neutron scattering\cite{McClarty_altermagnetism}.
We choose to study the AFM spin wave dispersion at the bottom of the dispersion at $(H00)$ for $H=1$ rlu. We aim to perform constant \textbf{Q}-scans with different resolution tails by varying the instrumental setup. 

For the dispersion in MnF$_2$, there exist an analytical model\cite{Yamani_2010}, which we have implemented in McStas, see Ref. \cite{Schack2025} for more details. We perform simulations of MnF$_2$ using a typical doubly-focusing cold-neutron TAS model described in appendix \ref{app:McStas_model}. We add two sets of slits and vary their width equally in order to perform simulations as a function of ingoing and outgoing angular divergence, which effectively means varying the $Q$-resolution. For each slit setting, we perform a constant-Q cut at the gap positions, and for each setting the resolutions are calculated from the covariance matrices given in McStas. The method for calculating such are found in appendix \ref{app:covariance_matrix}, and the resolutions are shown in Table \ref{tab:resolution_McStas}.

We report the three $Q$-resolution components; the in scattering-plane principal major axis $dQ_{\mathrm{prin \, major}}$ and principal minor axis $dQ_{\mathrm{prin \, minor}}$, while $dQ_{\mathrm{vert}}$ is the uncorrelated resolution vertical to the scattering plane. From the table, the two coarse $Q$-resolutions $dQ_{\mathrm{prin \, major}} \approx dQ_{\mathrm{vert}}$, which we use as $\sigma_Q$ in our analytical function eq.~\eqref{eq:analytical}. The narrow in-plane resolution, $dQ_{\mathrm{prin \, minor}}$, varies from 1/2 to 3/4 of the size of the other two. It turns out that when the narrow resolution is below around half the size of the other two, 
the approximations of two broad and one narrow $Q$-resolution components hold. \\
The energy resolution $dE_{\mathrm{at \, gap}}$ is the intrinsic instrumental energy resolution at a specific \textbf{Q}, namely at the gap position (orange in figure \ref{fig:resolution_assumptions} right). The last column is the projected energy resolution $dE_{\mathrm{proj}}$, green in figure \ref{fig:resolution_assumptions} right, which almost stays constant over all slit widths. This is to be expected, since we have a Rowland focused (energy focused) monochromator and analyser. The energy resolution $dE_{\mathrm{at \, gap}}$ is as expected smaller than the projected energy resolution, but opposite to the projected, it increases with increasing slit width. Which energy resolution to use, depends on the steepness of the dispersion as illustrated in the figure.

\begin{table}[ht!]
\centering
\caption{Simulated resolution FWHM components as a function of the slits width before and after the sample. The in-plane principal axes of the $Q$-resolution is given by $dQ_{\mathrm{prin \, major}}$ and $dQ_{\mathrm{prin \, minor}}$, while $dQ_{\mathrm{vert}}$ is the uncorrelated resolution vertical to the scattering plane. The energy resolution $dE_{\mathrm{at \, gap}}$ is the intrinsic instrumental energy resolution at a specific \textbf{Q}, namely at the gap position (orange in figure \ref{fig:resolution_assumptions} right). The last column is the projected energy resolution $dE_{\mathrm{proj}}$, green in figure \ref{fig:resolution_assumptions} right. The energy resolution is given in meV and the $Q$-resolutions are in units of Å$^{-1}$.}
\begin{tabular}{c c c c c c}
\hline
Slit width [cm] & $dQ_{\mathrm{prin \, major}}$ & $dQ_{\mathrm{prin \, minor}}$ & $dQ_{\mathrm{vert}}$ & $dE_{\mathrm{at \, gap}}$ & $dE_{\mathrm{proj}}$\\
\hline
0.5 & 0.0200 & 0.0148 & 0.0238 & 0.0259 & 0.1300 \\
1.0 & 0.0213 & 0.0196 & 0.0283 & 0.0479 & 0.1295 \\
2.0 & 0.0338 & 0.0255 & 0.0343 & 0.0692 & 0.1352 \\
3.0 & 0.0469 & 0.0301 & 0.0464 & 0.0870 & 0.1344 \\
4.0 & 0.0589 & 0.0384 & 0.0551 & 0.0925 & 0.1327 \\
5.0 & 0.0733 & 0.0427 & 0.0665 & 0.1010 & 0.1381 \\
6.0 & 0.0903 & 0.0470 & 0.0754 & 0.1059 & 0.1365 \\
7.0 & 0.0976 & 0.0540 & 0.0856 & 0.1061 & 0.1421 \\
8.0 & 0.1057 & 0.0551 & 0.0990 & 0.1057 &  0.1464 \\
\hline
\end{tabular}
\label{tab:resolution_McStas}
\end{table}


\subsection{Fitting to the data}
The simulated constant Q-cuts at increasing slit sizes are shown in figure \ref{fig:all_Mcstas_fits} in appendix \ref{app:fitting_param}. An example of the 5 cm slit width is shown in figure \ref{fig:McStats_fits}a. Here a clear peak is observed, with a resolution tail toward high energies.
In the simulations, no background scattering is simulated. Hence, to prevent zero-intensity points from being assigned unrealistically small uncertainties, the total intensity error for the fitting is defined for all points as the quadrature sum of the simulated error, $I_{err}$, and a constant term corresponding to 1\% of the maximum peak intensity, $I_{max}$; $\sqrt{I_{err}^2 + (0.01 \times I_{max})^2}$. For the fitting, we use the resolutions from Table \ref{tab:resolution_McStas}, namely $\sigma_E = dE$ and $\sigma_Q = \frac{1}{2}(dQ_{\mathrm{prin \, major}} +dQ_{\mathrm{vert}})$, as the initial parameters. Given the analytical expression for the dispersion in Ref. \cite{Nikotin_1969}, the slope of the dispersion is found to be $a=15.136$ meV/Å$^{-1}$. In the fits, the spin wave velocity, $a$, is fixed, while all other parameters are allowed to vary within the convoluted gap function in eq.~\ref{eq:analytical}. Since $a$ and $\sigma_Q$ are multiplied together, one needs to be kept fixed for fitting. For comparison, we also fit with a Gaussian line-shape and a combined Gaussian plus Voigt function (see appendix \ref{app:Gauss+Voigt}). 

\subsection{The fitted gap sizes}
\begin{figure}[ht!]
     \centering
     \begin{subfigure}[t]{0.65\textwidth}
         \centering
         \includegraphics[width=\textwidth]{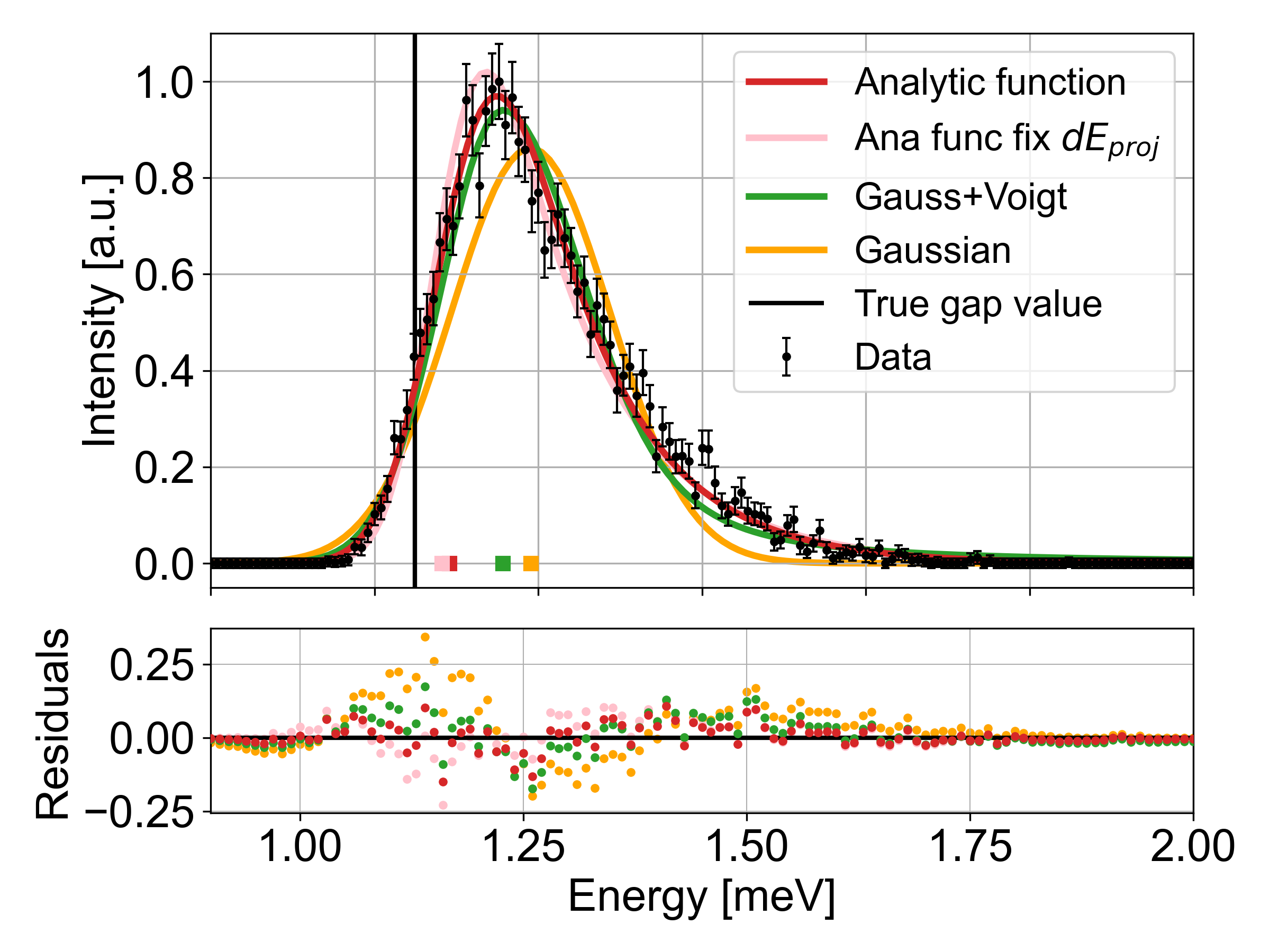}
         \caption{Example of simulated constant-\textbf{Q} scan through the spin wave gap in the McStas TAS model with slit of 5 cm. }
     \end{subfigure}
     \hfill
     \begin{subfigure}[t]{0.49\textwidth}
         \centering
         \includegraphics[width=\textwidth]{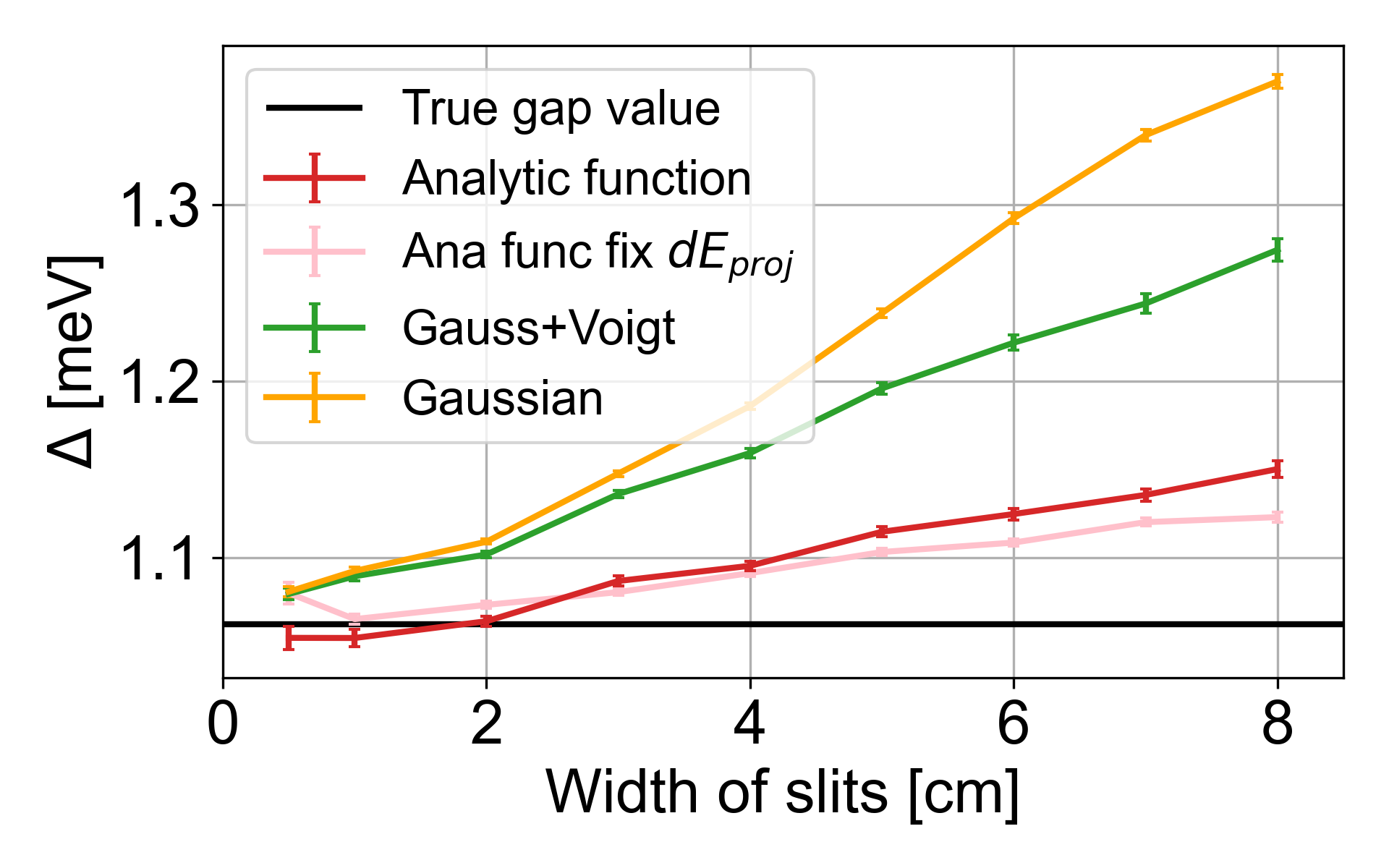}
         \caption{Gap size $\Delta$}
     \end{subfigure}
          \hfill
     \begin{subfigure}[t]{0.49\textwidth}
         \centering
         \includegraphics[width=\textwidth]{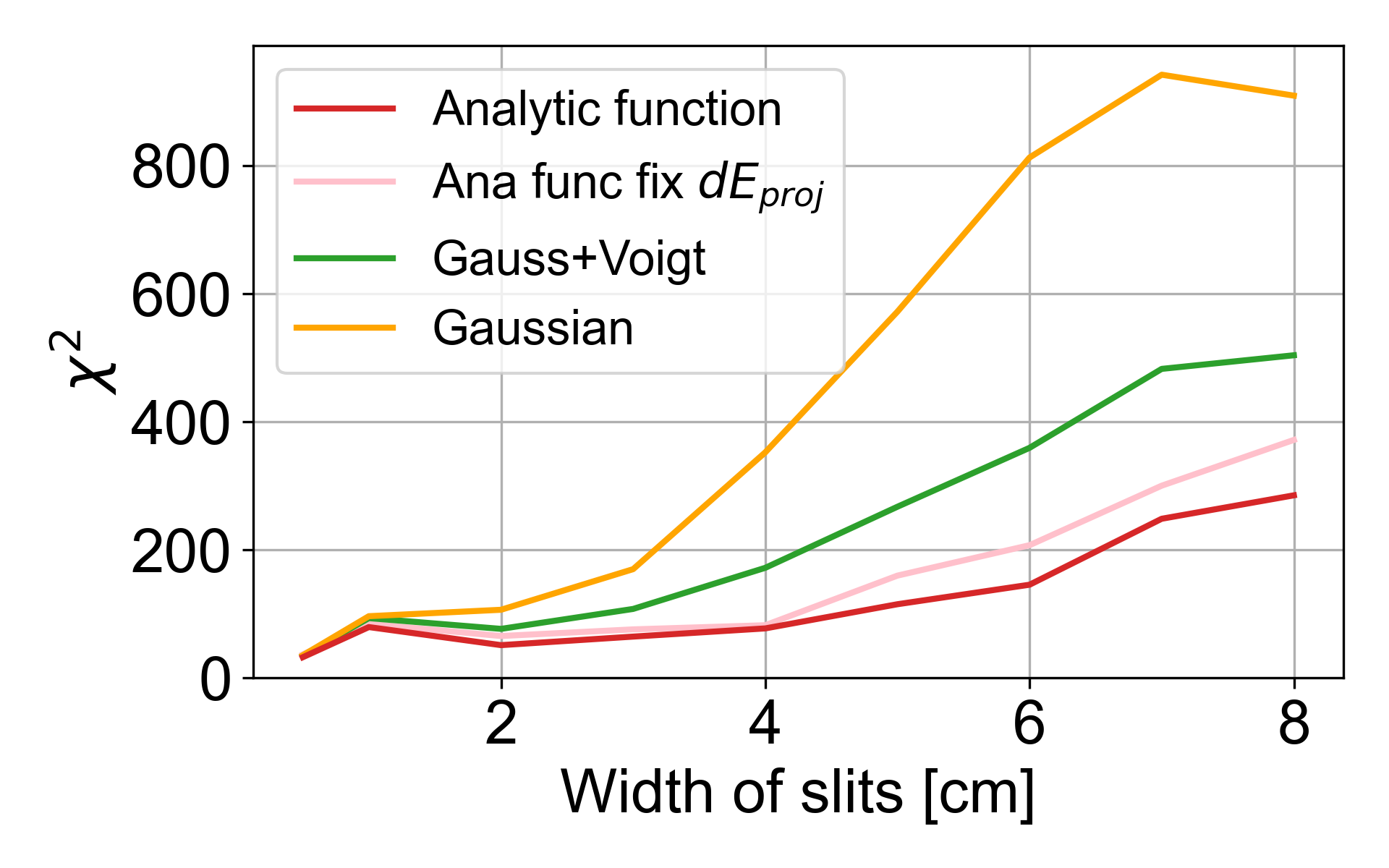}
         \caption{$\chi^2$ fits}
     \end{subfigure}
        \caption{a) The simulated data is plotted in black errorbars and is normalised to unity. The fitted peak positions are plotted at zero in coloured squares. The simulated MnF$_2$ data with a Gaussian (yellow line), a combined Gaussian, Voigt function (green line) and with our approximate convoluted-gap function (red line). The pink line shows a fit to the convoluted gap function with a fixed $\sigma_E$. The fitted gap positions are the coloured squares at zero and the black line shows the nominal gap size at (100), $\Delta_T=1.062$~meV. The residuals (data minus fit) are plotted below. The same colour coding is used in b) and c), showing the obtained gap values and $\chi^2$ values as a function of slit settings, respectively.
        }
        \label{fig:McStats_fits}
\end{figure}

From the three fits to the data from the 5~cm slit setting, figure \ref{fig:McStats_fits}a, the fitted gap sizes (coloured squares) are $1.239(2)$~meV (Gaussian), $1.196(3)$~meV (Gaussian plus Voigt) and $1.114(3)$~meV (analytic covoluted-gap function). From the input parameter to our analytical McStas model \cite{Yamani_2010, Schack2025} we obtain a nominal gap size at (100) of $\Delta_T=1.062$ meV (black, horizontal line in figure \ref{fig:McStats_fits}b). All functions overestimate the size of the gap; by 17\%, 13\% and 5\%, respectively. The $\chi^2$ value, which quantifies how well a model describes the data, are respectively 572.0, 267.5 and 115.3. It is clear that the Gaussian line-shape is not a good fit, while the combined Gaussian, Voigt function and our convoluted-gap function look to describe the data much better. Thus, from the example, our analytic gap-convoluted function gives a closer estimate of the gap and describes the data much better, than the other two functions.

Generally, looking across various slit widths, the same picture is presented. In figure \ref{fig:McStats_fits}b, the fitted gap sizes for all three fits are plotted as a function of the width of the slits, and figure \ref{fig:McStats_fits}c shows their respective $\chi^2$ values. From this, it is clear that our convoluted-gap function performs much better than the other two functions. However, as the resolution becomes larger, our function also tends to overestimate the gap size. 


\subsection{The fitted resolutions}
\begin{figure}[ht!]
     \centering
     \begin{subfigure}[t]{0.49\textwidth}
         \centering
         \includegraphics[width=\textwidth]{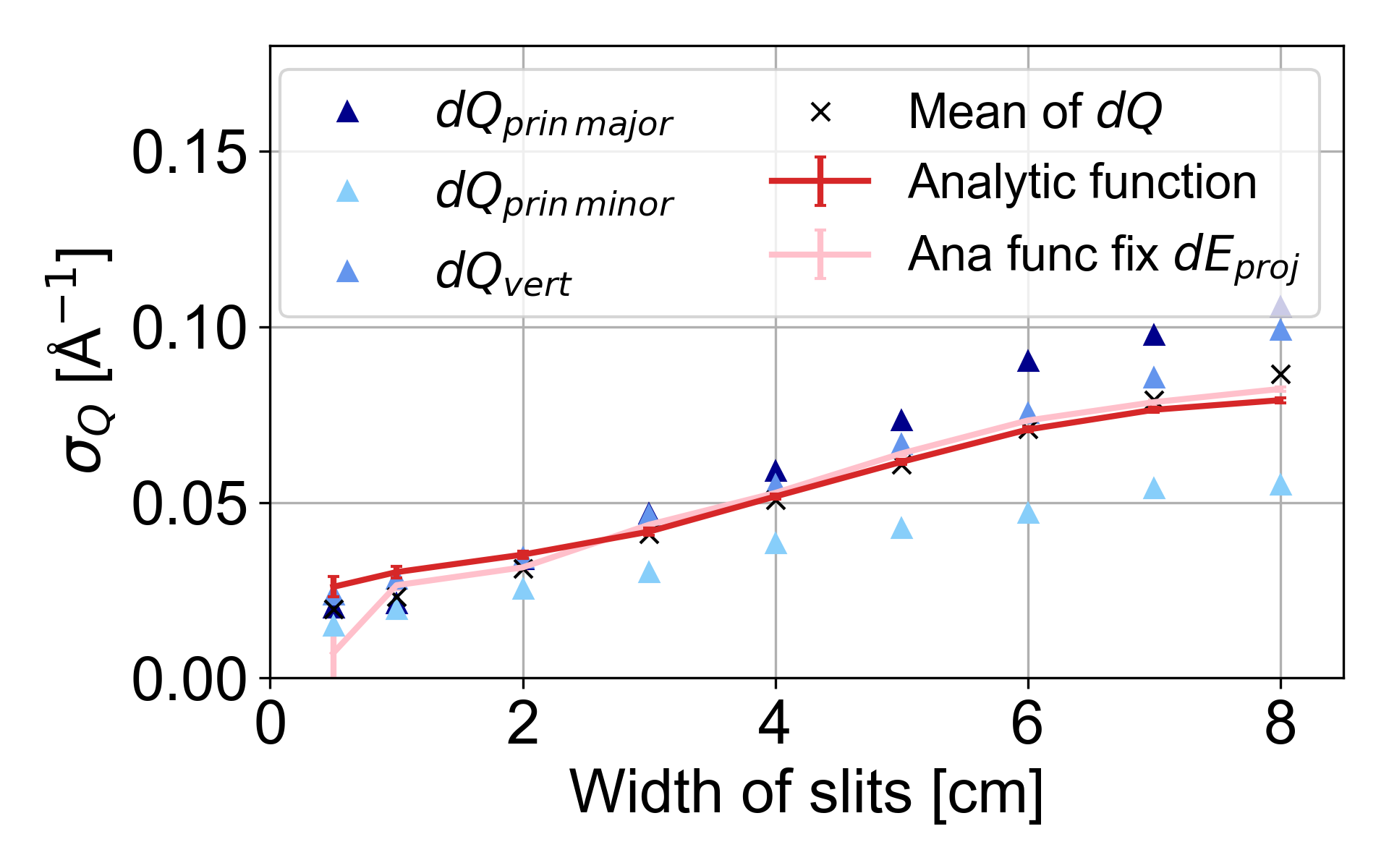}
         \caption{$Q$-resolution; $\sigma_Q$}
     \end{subfigure}
          \hfill
     \begin{subfigure}[t]{0.49\textwidth}
         \centering
         \includegraphics[width=\textwidth]{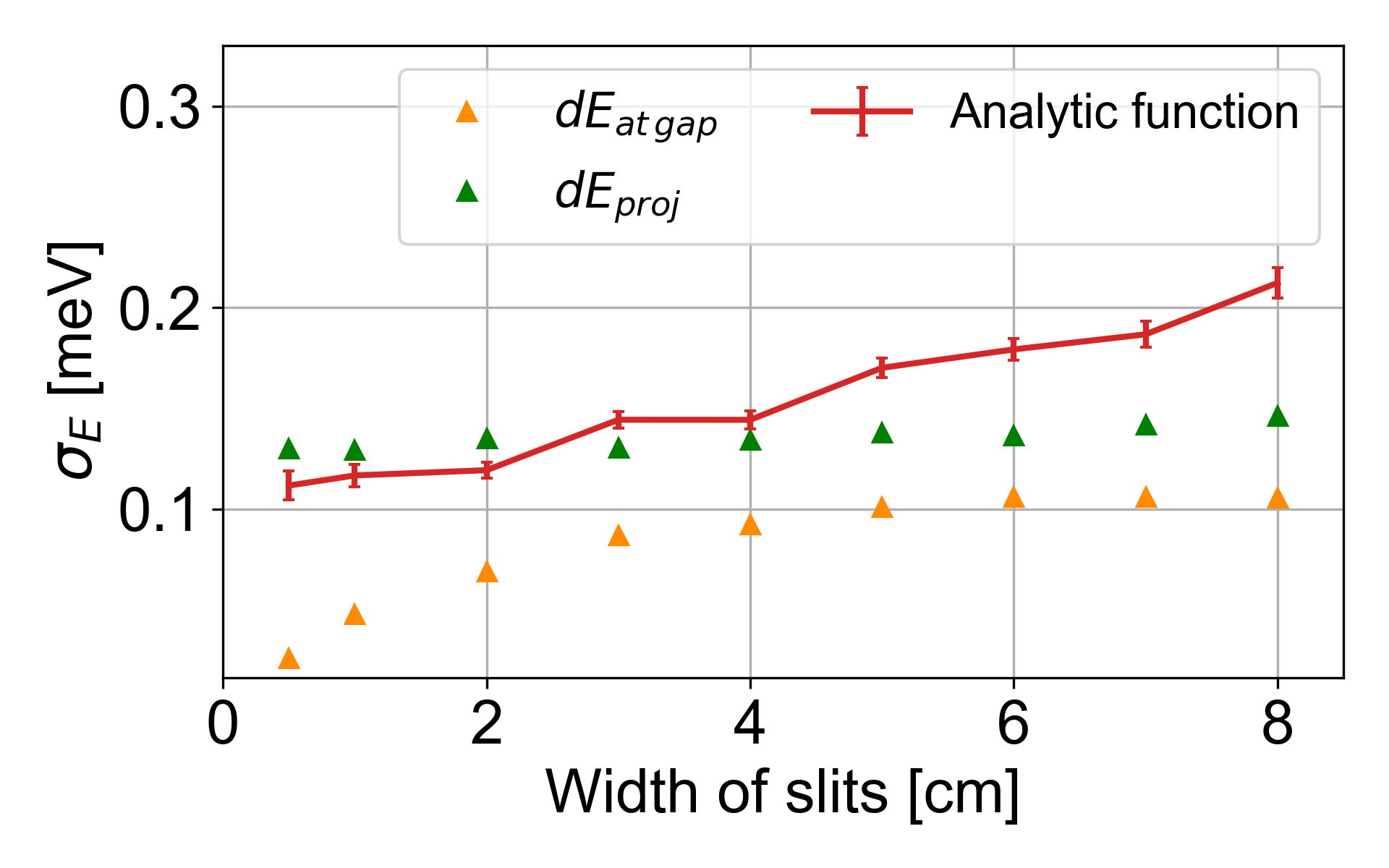}
         \caption{Energy resolution; $\sigma_E$}
     \end{subfigure}
        \caption{Fitted $Q$-resolution and energy resolution from our approximate convoluted gap function eq. \ref{eq:analytical}. The red line is where only $a$ is fixed in the fit, where in pink is both $a$ and $\sigma_E$ fixed. The triangles, in both figures, indicate the respective calculated resolutions from Table \ref{tab:resolution_McStas}. The black crosses in a) are the three calculated $Q$-resolutions averaged. All resolutions are FWHM.}
        \label{fig:McStats_fit_resolution}
\end{figure}

In the example of the 5 cm slits, the simulated resolutions are $\sigma_Q=0.0699$ Å$^{-1}$ (average of $dQ_{\mathrm{prin \, major}}$ and $dQ_{\mathrm{vert}}$) and the energy resolution is either the intrinsic at the gap $\sigma_E=0.101$ or the projected 0.138~meV. Given the steepness of the dispersion and the size of the resolutions, we have the broad dispersion case (figure \ref{fig:resolution_assumptions} right), and will work with the projected energy resolution. From the convoluted-gap function, the fitted parameters are $\sigma_Q=0.0615(7)$~Å$^{-1}$ and $\sigma_E = 0.170(5)$~meV. Thus, the $Q$-resolution is slightly underestimated by the fit, while the energy resolution is overestimated. The general trend for the fitted resolutions with the analytic function (eq.~\eqref{eq:analytical}) is plotted in figure \ref{fig:McStats_fit_resolution}, with the calculated resolutions from Table \ref{tab:resolution_McStas} plotted as triangles. Figure \ref{fig:McStats_fit_resolution}a shows the fitted FWHM $Q$-resolution (red errorbars) as a function of the slit widths, compared to the calculated resolutions from the covariance matrix (blue triangles). One would expect the fitted $Q$-resolution to be around the average of $dQ_{\mathrm{prin \, major}}$ and $dQ_{\mathrm{vert}}$ (dark blue triangles). At small slit widths, we observe the fitted $Q$-resolution is slightly overestimated compared to the expected values, while at 3 cm slits, it starts becoming slightly underestimated. Taking instead the mean of all three $Q$-resolutions (black crosses), we obtain a rather accurate estimate of the fitted $Q$-resolution. This is probably due to the fact that our resolution is not a real pancake, meaning that $dQ_{\mathrm{prin \, minor}}$ is not a delta-function. \\
For the fitted FWHM energy resolution in \ref{fig:McStats_fit_resolution}b (red errorbars), a slight increase is observed with increasing slit width. The intrinsic resolution at the gap, $dE_{\mathrm{at \, gap}}$ (orange triangle), is smaller than the fitted resolutions, indicating that our dispersion is broad (see figure \ref{fig:resolution_assumptions} right), such that we need to compare to the projected energy resolution, $dE_{\mathrm{proj}}$ (green triangle). Comparing the fitted energy resolutions to the projected, opposite the $Q$-resolution, at small slit widths the fitted resolution is slightly underestimated, while at 3 cm slit width, it becomes overestimated. Thus, it seems that the fitting function at large slit widths underestimates the $Q$-resolution ($12\%$) and compensates by overestimating the energy resolution ($23\%$ compared the projected energy resolution).\\

\subsection{Estimation of the high-energy tail}
Since we know the "true" resolutions, from table \ref{tab:resolution_McStas}, we have performed fits with fixed value of the energy resolution, $\sigma_E = dE_{\mathrm{proj}}$. These fits are all plotted in pink in figure \ref{fig:McStats_fits} and figure \ref{fig:McStats_fit_resolution}. This procedure gives a gap size slightly closer to the true gap value; with only a slightly higher value of $\chi^2$. 

\begin{figure}
    \centering
    \includegraphics[width=0.65\textwidth]{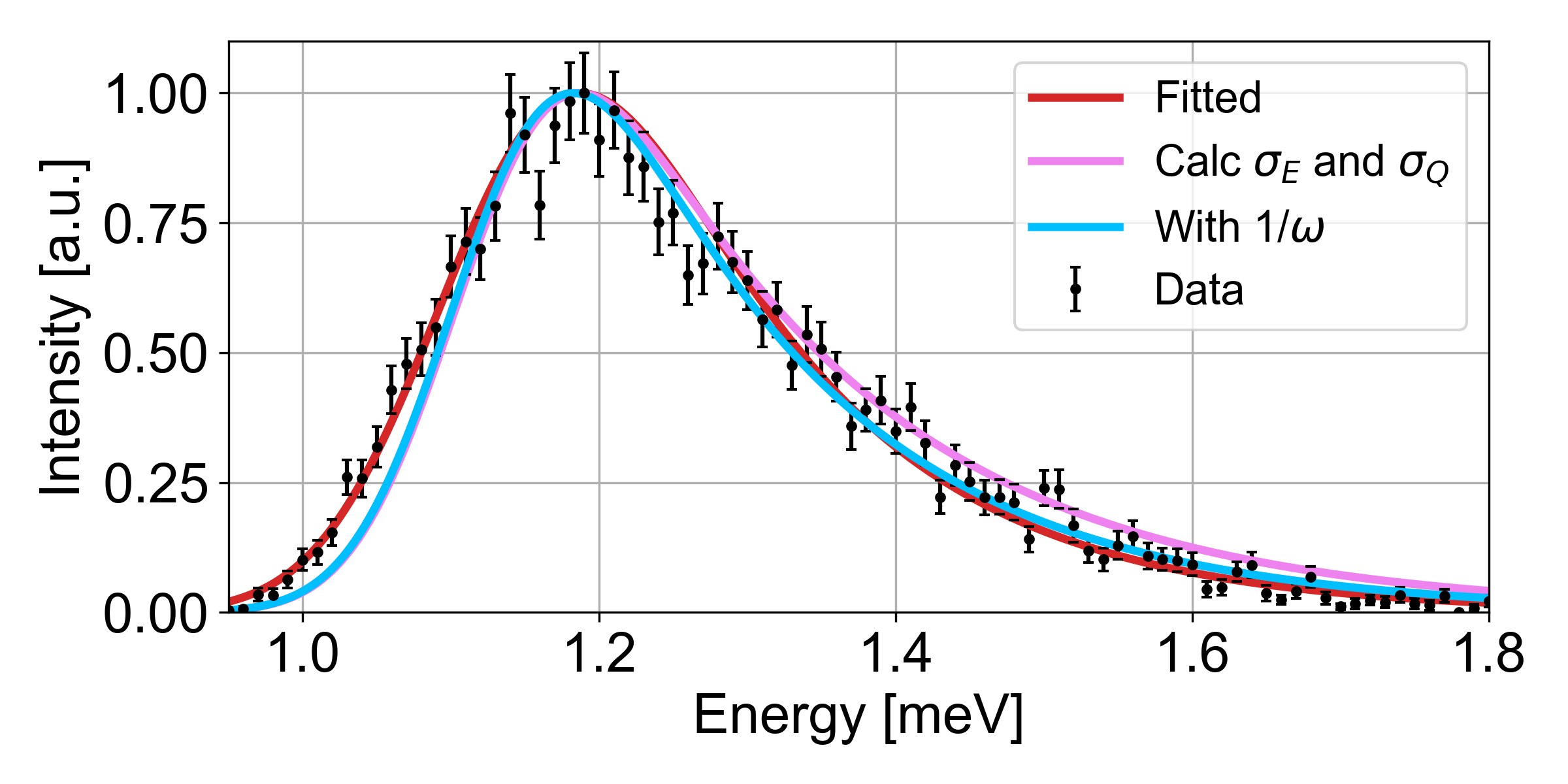}
    \caption{Simulated constant-\textbf{Q} scan through the spin wave gap with slit of 5 cm, data shown in black errorbars and the fitted approximate convoluted-gap function is plotted in red, same as figure \ref{fig:McStats_fits}a. Replacing the resolutions by the calculated McStas $\sigma_E$ and $\sigma_Q$, Table \ref{tab:resolution_McStas}, are plotted in violet. Eq. \eqref{eq:numerical_omega} is plotted in blue, here the calculated McStas resolutions are used.
    The parameters are the same for all plots, except if otherwise mentioned.}
    \label{fig:resolution_omega}
\end{figure}

As noted above, to fit the tail the routine underestimates the $Q$-resolution and compensates by overestimating the energy resolution, slightly inflating the extracted gap size. In the 5 cm slit width case, using the calculated rather than fitted values for $\sigma_Q$ and $\sigma_E$ makes the tail larger, see figure \ref{fig:resolution_omega}.
Here we present a zoom in on the 5 cm slits scan with the fitted parameters of the analytic convoluted-gap function in eq.~\eqref{eq:analytical} (same as figure \ref{fig:McStats_fits}a). To account for this effect, we have to look at the approximations behind our analytic convoluted-gap function; the AFM magnon dispersion at the gap position can be described by a parabola and that $S_n(\textbf{Q})$ in the structure factor (eq.~\eqref{eq:S(Q,omega}) is constant. If the resolution is large such that we pick up intensity from the parobola sides, both approximations will contribute to an overestimation of the tail. To go beyond these assumptions, we include $S_n(\textbf{Q})\propto 1/\omega$ in eq.~\eqref{eq:delta_func_approx}, such that eq.~\eqref{eq:fitting_func} becomes: 
\begin{equation}
        I(\mathbf{Q}, \omega) \propto A_1 \int \frac{1}{\omega'}f(\omega') \exp \left( - \frac{(\hbar\omega - \hbar\omega')^2}{2 \sigma_E^2} \right) \, d\omega' + B.
        \label{eq:numerical_omega}
\end{equation}
Here $A_1$ includes all constants. The integral in this expression must be computed numerically, which makes it unstable in fitting procedures. In figure \ref{fig:resolution_omega}, we have plotted eq. \eqref{eq:numerical_omega} with the calculated McStas resolutions; $\sigma_Q$ and $\sigma_E$. From this, we see that with the $1/\omega$ factor, the tail is only slightly reduced, therefore, we deem it unimportant to include the factor. Testing the other approximation, the approximation of the parabola in eq. \eqref{eq:parabola}, we find deviations when the resolution ellipsoid hits the linear part of the dispersion, see figure \ref{fig:error_region} in appendix \ref{app:limitations_parab}. However, in our fits the error on the gap size is relatively small ($5\%$ for 5 cm slit width). This is likely caused by the fact that the fits chooses a smaller $\sigma_Q$ and a larger $\sigma_E$ than the correct values. However, in instruments with a broader Q-resolution, deviations are likely to be more severe. 

\section{Experimental data; \texorpdfstring{MnF$_2$}{MnF2}}
\subsection{Experimental set-up}
Experimental data on MnF$_2$ were taken at the TAS-like neutron spectrometer CAMEA at the facility SINQ at the Paul Scherrer Institut (CH). This instrument facilitates a large analyzer-detector array in a vacuum tank, giving quasi-continuous coverage along energy transfer and scattering angle. By scanning the sample rotation, one can obtain coverage in two dimensions in Q-space spanned by two reciprocal lattice vectors (here [H00]-[00L]), and energy \cite{Lass2023}. 

The 3.446(1)~gram MnF$_2$ sample was positioned in an orange cryostat, and data were acquired at a temperature of 10~K by performing sample rotation scans for 5 different incoming neutron energies, ($E_{\rm i} = 5$, 5.5, 7, 8.5 and 10~meV), each using two different angles of the tank to cover dark angles. The data were converted and analysed using the dedicated software package MJOLNIR (version 1.3.1.post4) \cite{Lass2020MJOLNIR,jakob_lass_2023_8183140}.

\subsection{Experimental results}
\begin{figure}
    \centering
    \includegraphics[width=0.5\linewidth]{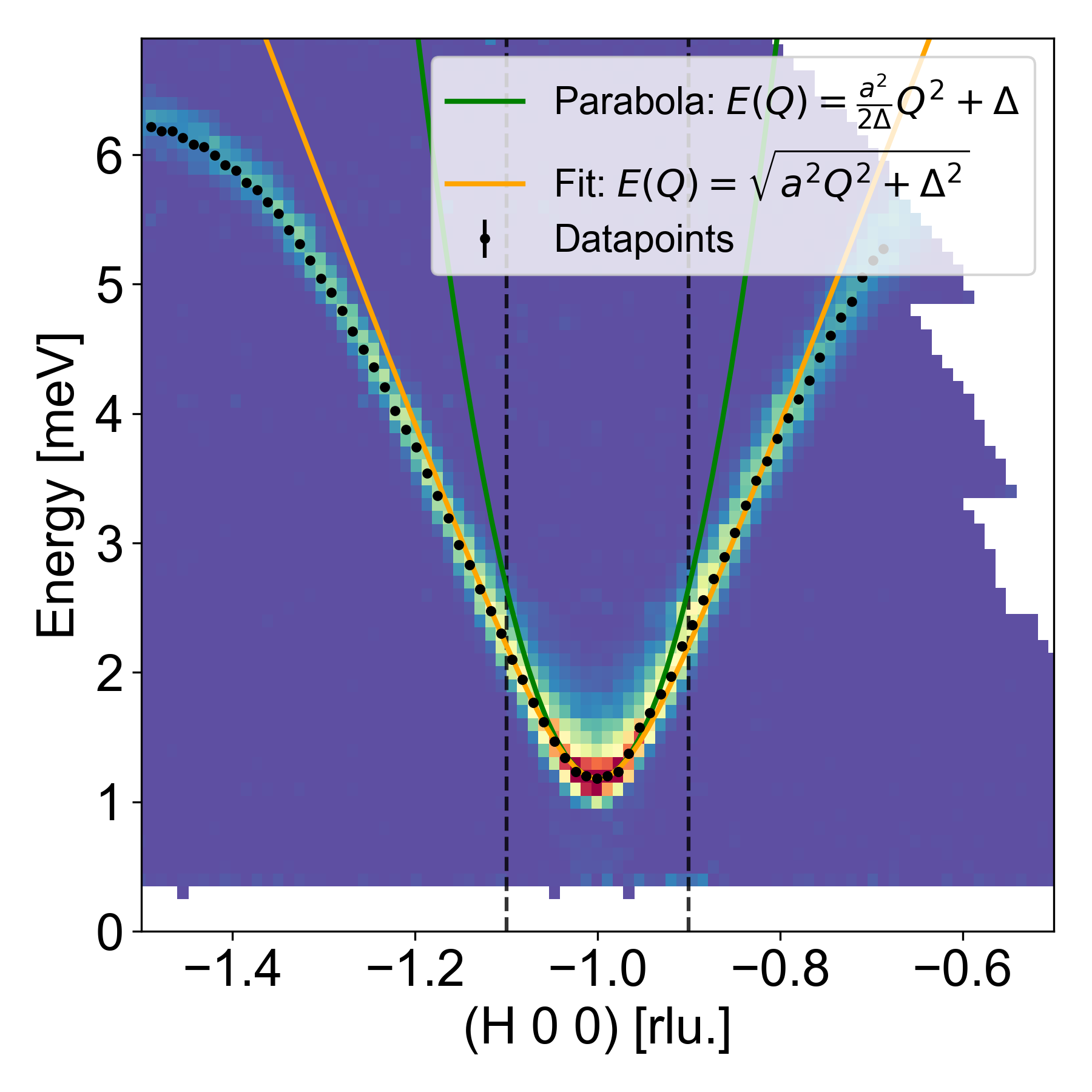}
    \caption{Dispersion of AFM MnF$_2$ along (H00) measured on the TAS-like instrument CAMEA (PSI) at 10 K. The black errorbars are constant Q-cuts fitted with a Gaussian line-shape to find the mode position. Eq. \eqref{eq:AFM_dispersion} is fitted to the errorbars in the dashed range (in orange) to find the spin wave velocity, $a=14.89 \pm 0.02$ meV/Å$^{-1}$. In dark green, is plotted the parabola approximation (eq. \eqref{eq:parabola}).}
    \label{fig:CAMEA_data}
\end{figure}

We choose to study the AFM spin wave dispersion along $(H00)$ for $-1.5<H<-0.5$, where the bottom of the dispersion is at $H=-1$, see figure \ref{fig:CAMEA_data}. Performing constant \textbf{Q}-cuts along $E$ and fitting the mode position in each cut with a Gaussian line shape, gives the data shown as black errorbars. These points are, in turn, used to fit the AFM dispersion eq. \eqref{eq:AFM_dispersion} in the linear range (within the dashed vertical lines) to find the slope of the dispersion, $a=14.89 \pm 0.02$ meV/Å$^{-1}$. This is very similar to the value found from the analytical expression mentioned in the previous section. 
The resolutions of the CAMEA instrument are calculated in MJOLNIR, which in turn uses the functionality of Takin\cite{Takin2023}. From this, we obtain the covariance matrix and can in turn calculate the principal axes. CAMEA measures the gap position, $\hbar \omega \approx 1.1$~meV, with two incoming energies; $E_{\rm i} = 5$~meV and 5.5~meV, for which we can calculate the resolution ellipses. The $Q$-resolution (FWHM) of the principal axes are the same for $E_i=5$ meV and 5.5 meV, the major is 0.116 Å$^{-1}$ and minor is 0.021 Å$^{-1}$. Here, the out-of-plane $Q$-resolution (0.095 Å$^{-1}$) is almost the same size as the major principal axis, thus we have two broad and one narrow $Q$-resolution component as prescribed by our approximations.
The energy resolution of CAMEA at the gap position has FWHM for $E_{\rm i}=5$~meV an intrinsic energy resolution of 0.087~meV and projected 0.185 meV. While the $E_i=5.5$ meV is slightly different, with an intrinsic energy resolution of 0.090~meV and projected 0.213~meV. Both ellipses are plotted in figure \ref{fig:CAMEA_data_fits}b. These resolutions are broader than in the simulated TAS instrument above.

The spin wave spectrum of MnF$_2$ has an experimentally determined gap size of 1.081 meV measured through inelastic neutron scattering \cite{Nikotin_1969, Low_1964} and antiferromagnetic resonance (AFMR) experiments\cite{Johnson_1959}. AFMR probes the $\textbf{Q}=0$ (uniform) magnon modes of an AFM by measuring their microwave or terahertz resonance frequencies. Since, it directly measures these zero-momentum excitations, AFMR can determine the spin gap with exceptionally high energy precision; typically on the order of $\mu$eV.

\subsection{Fitting the experimental data}
\begin{figure}[ht!]
     \centering
     \begin{subfigure}[t]{0.6\textwidth}
         \centering
         \includegraphics[width=\textwidth]{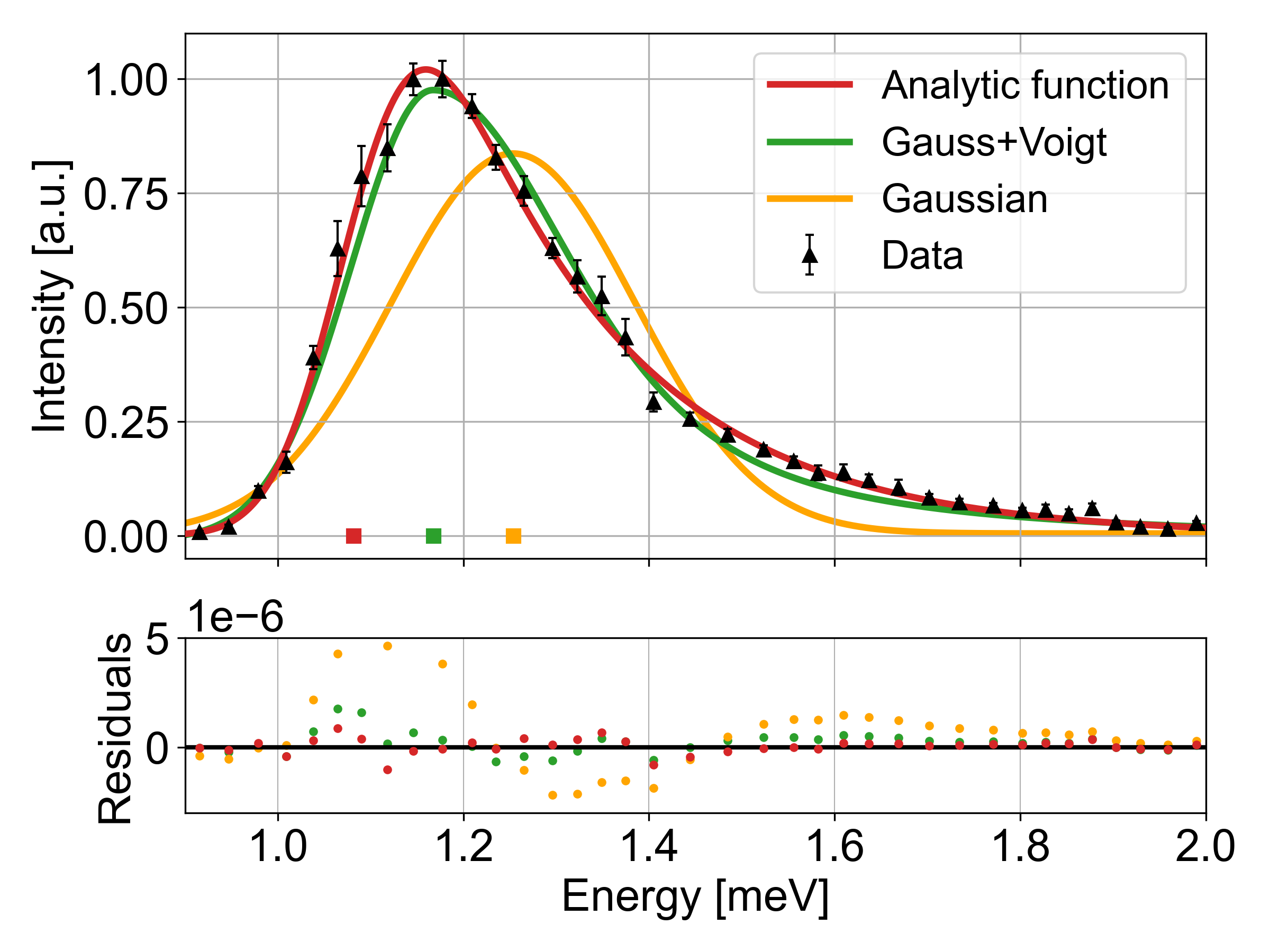}
         \caption{ }
     \end{subfigure}
     \hfill
     \begin{subfigure}[t]{0.33\textwidth}
         \centering
         \includegraphics[width=\textwidth]{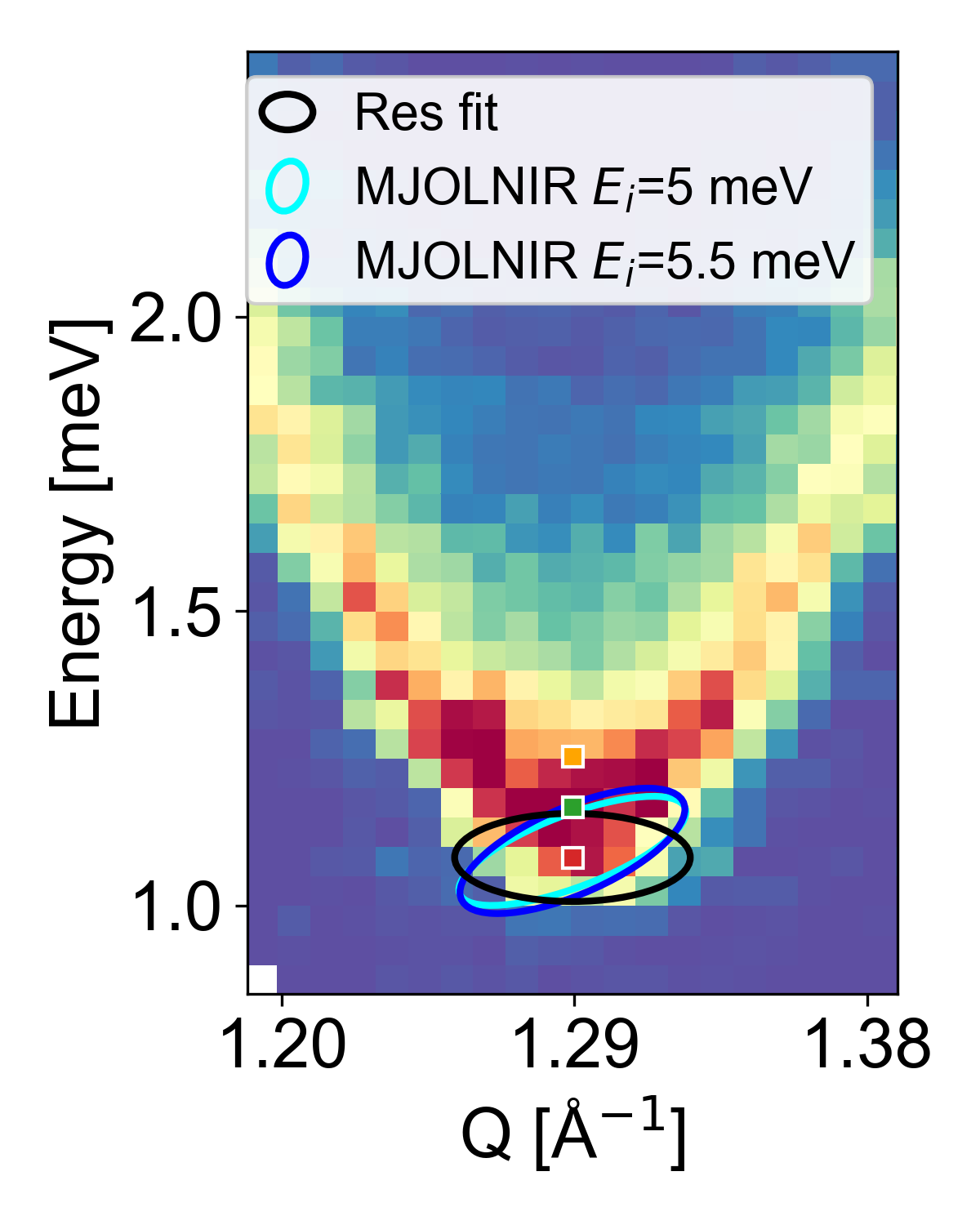}
         \caption{ }
     \end{subfigure}
        \caption{Fitting the gap size of the CAMEA data 10 K with a Gaussian (yellow line), a combined Gaussian, Voigt in eq. \ref{eq:Gauss_Voigt} (green line) and with our approximate convoluted-gap function eq. \ref{eq:analytical} (red line). The fitted gap sizes are plotted in coloured squares. a) Constant Q-cut at (H00) H$=-1$ with an integration width of 0.02 rlu. The residuals (data minus fit) are plotted below. b) Zoom in of the dispersion in figure \ref{fig:CAMEA_data} with the fitted gap sizes from a). The black ellipsis indicates the fitted resolution, and the two blue ellipses are the onces calculated in MJOLNIR. }
        \label{fig:CAMEA_data_fits}
\end{figure}

We perform a constant-Q cut at (H00) H$=-1$ with an integration width of 0.02 rlu. in both $Q$-directions in the scattering plane, see figure \ref{fig:CAMEA_data_fits}a. For the fit, we keep $a$ fixed and all other parameters are fitted with eq. \eqref{eq:analytical}. Again, we compare our function to a Gaussian line-shape and a Gaussian plus Voigt function (see appendix \ref{app:Gauss+Voigt}). From the figure it is clear that the Gaussian provides a bad fit of the gap data; it finds the gap value to be 1.254(2)~meV, which is highly overestimated. The Gauss plus Voigt function and the convoluted-gap function both describe the data much better. The fits yield gap sizes of 1.167(3)~meV and 1.082(2)~meV, respectively, with $\chi^2$-values of 502.3 and 200.7, respectively. The fitted gap values are plotted on top of the dispersion in figure \ref{fig:CAMEA_data_fits}b. From the excellent agreement with the gap value measured by AFMR, it is clear that our analytical function provides a very accurate means of determining the gap value from the neutron data. \\
The fitted resolutions are $\sigma_E=0.149(4)$~meV and $\sigma_Q=0.0725(5)$~Å$^{-1}$. Compared to the calculated resolutions, $\sigma_E$ is slightly underestimated in respect to the projected resolution. On the other hand, $\sigma_Q$ is also underestimated compared to the major principal axis. However, the average of the $Q$-resolutions in the three directions is 0.0773~Å$^{-1}$, which is very similar to the fitted value, as we also observed from the fits to the simulated data. 

One can also decide to fix the energy resolutions to the calculated values. We have performed this on our CAMEA data, using both the projected energy resolution and the intrinsic energy resolution at the gap. The fits are shown in appendix \ref{app:CAMEA_fit_fix_dE}. From these fits we obtain gap values of 1.097(2)~meV and 1.066(2)~meV with $\chi^2$-values of 277.2 and 395.2, respectively. Thus, fixing the energy resolution to the projected energy resolution (0.185 meV), we get a slightly worse fit, but the fitted gap size becomes larger than the previously reported gap size. In analogy, fixing the energy resolution to the intrinsic energy resolution at the gap (0.087~meV), we also obtain a non-ideal fit. In the latter case, however, we force the gap to be smaller due to the fit mainly takes the high-energy tail into account. Thus, the obtained fitted energy resolution (0.149 meV) takes a value in between.

\section{Discussion and conclusion}
In this manuscript, we have presented an analytic convoluted-gap function, eq.~\eqref{eq:analytical}, which is based on two  approximations: (1) the $Q$-resolution is coarse in two out of three directions (typical of a focused TAS) and (2) at the gap, the dispersion follows a parabola. For quadratic excitations, such as FM spin waves, the parabola is an exact representation, while for linear excitations, such as AFM spin waves, it is merely an approximation. We tested the approximation on AFM spin waves in MnF$_2$ on simulated data, while effectively varying the $Q$-resolution. We also tested with experimental data from the TAS-like instrument CAMEA in a single experimental setting. In both cases, the convoluted-gap function performs much better than a Gaussian line-shape or a Gaussian plus Voigt function.


From our simulated data we find that with increasing $Q$-resolution (slit width), the fitting routine tends to underestimate $\sigma_Q$, compared to what was expected from the mean of the two coarse $Q$-resolutions. Instead, taking the average of the three calculated $Q$-resolutions, one find very good agreement with the fitted $Q$-resolution. For the underestimated $\sigma_Q$, the fitting routine compensates with increasing $\sigma_E$, such that the gap size $\Delta$ in effect becomes slightly overestimated. We ascribe this to our parabolic approximation for the bottom of the AFM dispersion, which in turn overestimates the high-energy tail above the gap.

We tested our assumption of a constant $S_n(\textbf{Q})$ by in stead using the more realistic scenario $S_n(\textbf{Q}) \propto 1/\omega$.
This results in a very slight overestimation of the high-energy tail, and a small narrowing of the low-energy part, which might result in a slight reduction of the gap size. However, the resulting function must be evaluate numerically and struggles to converge when implementing it into a fitting routine. The overestimation of the high-energy tail is minor, and our analytic function compensates for it by adjusting $\sigma_Q$
when it is kept as a free fitting parameter. 

For the TAS-like experimental data, the analytic function performs very well, even though CAMEA is a much more complicated instrument than a standard TAS. Thus, our function hs proven in practice to be a good approximation for fitting gaps, where the resolution function is much narrower than the gap value.

We find that the parabolic approximation provides an excellent local model at the gap and for small displacements from the minimum, capturing the essential curvature of the linear excitation response with errors that remain below a few percent. However, it should not be extended too far into the wings of the function, where the difference grows rapidly, and the true linear excitation departs significantly from the simple quadratic form. 

As a further development, it is possible to extend the expression to that of a double gap. This was done by us in Ref. \cite{Lenander_2025}, where we found that the function works well for fitting the size of the gaps, while the function struggles to get the resolution tails of the two gaps correctly. 
Another use case is to investigate whether a 2-magnon scattering continuum is present. To determine whether the high-energy tail contains continuum scattering, the resolution has to be well known. Using the correct resolutions (like figure \ref{fig:resolution_omega} in purple), if the data points lie above the analytical function, one can be certain that extra scattering is present and contribute to the high-energy tail. While, if the data lies below the function, then a more thorough investigation is needed. We also attempted to determine a continuum of scattering in Ref. \cite{Lenander_2025}, however, this was inconclusive due to the presence of a double gap.

To conclude, we have derived an analytical expression for the resolution convoluted line shape of a scan through a gap. We have demonstrated, both on simulated and experimental data, that the analytical function provides a good fit to the data. As a result, it provides a significant better estimate of the gap value than using a Gaussian line-shape or a Gaussian plus Voigt function. The analytical function can easily be implemented into fitting routines, without suffering from the difficulties of convergence. 
Therefore, we recommend our analytical function for fitting of gaps values measured by TAS.

\begin{acknowledgements}
This work is partially based on experiments performed at the Swiss spallation neutron source SINQ, Paul Scherrer Institute, Villigen, Switzerland. We thank Jakob Lass for providing us with the CAMEA data of MnF$_2$.

We thank Kristine Krighaar for providing the generic Triple-Axis Spectrometer model in McStas. 

\end{acknowledgements}

\begin{funding}
This project was supported by the Danish Committee for Research Infrastructure (NUFI) through funding to the ''ESS-Lighthouse`` Q-MAT and the Swiss National Science Foundation (SNSF) project 200021-228473 - Quantum Magnetism and Spectroscopy.

\end{funding}

\ConflictsOfInterest{The authors declare no conflicts of interest.
}

\DataAvailability{The simulated data and the CAMEA data can be accuried from the authors upon request. 
}

\bibliography{iucr} 

\newpage
     
\appendix 

\section{Guide to experimentally determine the parameters in the analytic function}\label{app:scanning_Bragg_peak}
When performing a TAS experiment, one can determine the instrument resolution experimentally by performing a few scans; 
\begin{itemize}
  \item The in-plane $Q$-resolution; $dE_{\mathrm{prin \, major}}$ and $dE_{\mathrm{prin \, minor}}$. \\
  First perform a crystal rotation scan over the Bragg peak position, then extend the length of \textbf{Q} by 1\% and repeat. If the peak intensity is still above half of the central scan, then double the step size in the length of \textbf{Q} and perform a new crystal rotation scan. From this, you will find your in-plane $Q$-resolution ellipsis. 
  \item The out-of-plane $Q$-resolution; $dE_{\mathrm{vert}}$. \\
  This can be found by tilting the crystal around an in-plane axis perpendicular to the scattering plane. Then performing the same type of scan as mentioned above. 
  \item The energy resolution; $dE_{\mathrm{proj}}$ or $dE_{\mathrm{at \, gap}}$. \\
  Depending on the steepness of the dispersion at the gap compared to the resolution, one need to use either the projected energy resolution or the intrinsic energy resolution at the gap position (see figure \ref{fig:resolution_assumptions} right) in the approximation. For a broad dispersion one uses the projected energy resolution, which one commonly determines through a scan of an incoherent scatter, such as vanadium. For a narrow dispersion, the intrinsic energy resolution at the gap position is used and it can be found by performing an energy scan through the Bragg peak. 
  \item The spin wave velocity, $a$. \\
  By measuring a constant-energy scan at approximately two times the gap size, you can find the mode positions at the linear part of the dispersion and then use eq. \eqref{eq:AFM_dispersion} to fit $a$. 
  \item The curvature of the parabola, $\alpha$. \\
  The same approach as above, but use eq. \eqref{eq:parabola} instead.
\end{itemize}
Note, that this guide finds the $Q$-resolution and the energy resolution for elastic scattering, which can vary slightly to the resolution for inelastic scattering. However, it will provide an estimate of the instrumental resolution at the gap, assuming that the gap is close to the elastic line.

\section{Gaussian and Voigt functions}\label{app:Gauss+Voigt}
To fit a dispersion to find the mode positions, one will often select a Gaussian lineshape, defined as: 
 \begin{equation}
     f(x; A, \mu, \sigma) = A \exp \left ( -\frac{(x-\mu)^2}{2\sigma^2}\right )
     \label{eq:Gaussian}
 \end{equation}
where $A$ is the normalisation, $\mu$ is the centre of the peak and $\sigma$ is the standard deviation or width of the peak. 

However, at the bottom of a dispersion, you tend to pick up signal from higher energies, resulting in a tail at energies above the gap. To describe the tail, a Gaussian for the lower energy part of the peak and a Voigt function for the upper energy part of the peak can be used. The Voigt profile is a convolution of a Gaussian and a Lorentzian function, typically having a Gaussian peak shape with a Lorentzian tail. The Voigt profile is defined as:
\begin{equation}
V(x; \mu, \sigma, \gamma) = \frac{\Re\left[w\left( \frac{x - \mu + i \gamma}{\sigma \sqrt{2}} \right)\right]}{\sigma \sqrt{2\pi}}
\end{equation}
where $\mu$ is the centre of the peak, $\sigma$ is the standard deviation of the Gaussian component and $\gamma$ is the half-width at half-maximum of the Lorentzian component. Also, \( w(z) \) is the Faddeeva function\cite{Voigt_Faddeeva} and \( \Re \) denotes the real part. To make the peak asymmetric composed of a Gaussian at lower energies and a Voigt profile at higher energies, the peak is given:
\begin{equation}
f(x; A, \mu, \sigma, \gamma) = 
\begin{cases}
A \cdot \exp\left( -\frac{(x - \mu)^2}{2\sigma^2} \right), & \text{if } x < \mu \\
A \cdot N \cdot V(x; \mu, \sigma, \gamma), & \text{if } x \geq \mu
\end{cases}
\label{eq:Gauss_Voigt}
\end{equation}
The factor $N=\frac{\exp\left( -\frac{(x - \mu)^2}{2\sigma^2} \right)}{V(\mu; \mu, \sigma, \gamma)}$ ensures continuity at $x=\mu$ by scaling the Voigt profile to match the Gaussian at the peak centre.

\section{Derivation of the convoluted-gap function}\label{app:derivation}

\subsection{The numerical derivation}\label{app:derivation_numerical}
Using the approximations presented in the main text, namely eq. \eqref{eq:resolution_approx} and eq. \eqref{eq:delta_func_approx}, we can write the intensity given by eq. \eqref{eq:intensity} as:
\begin{eqnarray}
I(\mathbf{Q}, \omega)  &\propto & \int_{-\infty}^\infty \delta(Q_{x0} - Q_x') \exp \left( -\frac{ (Q_{y0} - Q_y')^2 + (Q_{0z} - Q_z')^2}{2 \sigma_Q^2} - \frac{\hbar^2(\omega - \omega')^2}{2 \sigma_E^2} \right) \nonumber \\
 & \times & \delta(\hbar\omega' - \Delta - \alpha (\textbf{Q}-\textbf{Q}_0)^2) \; d\mathbf{Q'} \, d\omega'.
\end{eqnarray}
First, to simplify matters, we assume the gapped dispersion to be centred around zero, such that $\textbf{Q}_0=\textbf{0}$. We use polar coordinates to rewrite $Q_y'^2+Q_z'^2=Q^2$ and $d\textbf{Q}'=QdQd\theta$,
\begin{equation}
    I(\mathbf{Q}, \omega) \propto \int_{-\infty}^\infty \exp \left( -\frac{Q^2}{2 \sigma_Q^2} - \frac{\hbar^2(\omega - \omega')^2}{2 \sigma_E^2} \right) \delta(\hbar\omega' - \Delta - \alpha Q^2) \, Q \, dQ \, d\theta \, d\omega'.
\end{equation}
Substituting, $Q^2=(\hbar\omega-\Delta)/\alpha$ to get $dQ=(2Q\alpha)^{-1}d\omega$, which is used to simplify the expression,
\begin{equation}
\begin{split}
    I(\mathbf{Q}, \omega) 
    & \propto \int_{-\infty}^\infty \exp \left( -\frac{|\hbar\omega' - \Delta|}{2 \alpha \sigma_Q^2} - \frac{(\hbar\omega - \hbar\omega')^2}{2 \sigma_E^2} \right) \cdot \frac{\pi}{\alpha} \, d\omega'. \\
\end{split}
\end{equation}

Importantly, we only start picking up intensity when we reach the gap position, $\hbar\omega>\Delta$, so we use the Heaviside step, $H$,
\begin{equation}
    f(\omega) = H(\hbar \omega - \Delta) \exp \left( -\frac{|\hbar\omega - \Delta|}{2\alpha \sigma_Q^2} \right) =
\begin{cases}
\exp \left( -\frac{|\hbar\omega - \Delta| }{2 \alpha \sigma_Q^2} \right), & \hbar\omega > \Delta \\
0, & \hbar\omega \leq \Delta
\end{cases}
\label{eq:heavyside_app}
\end{equation}
The intensity becomes an asymmetric function $f(\omega)$ that is being convoluted with the energy resolution. We add the Gaussian normalisation constants and for experimental data, all other constants e.g. the prefactor of $S(\textbf{Q}, \omega)$ and the intensity normalisation of the instrument are described by a normalisation constant, $A$. To account for the instrument background, the scalar, $B$, is used. Thus, the function for fitting a gapped dispersion becomes: 
\begin{equation}
    I(\mathbf{Q}, \omega) = A \frac{1}{\sqrt{2 \pi \sigma_E^2}} \frac{1}{2 \alpha \sigma_Q^2}\int_{-\infty}^\infty f(\omega') \exp \left( - \frac{(\hbar\omega - \hbar\omega')^2}{2 \sigma_E^2} \right) \, d\omega' + B.
    \label{eq:fitting_func_app}
\end{equation}
However, this is a function to be determined numerically, which makes it in practice more difficult for a fitting routine to converge at the global minimum. Thus, we are interested in finding an analytical expression.

\subsection{The analytical derivation}\label{app:derivation_analytical}
We define the integral in eq. \ref{eq:fitting_func_app} by $J$:
\begin{equation}
J \equiv \int_{\Delta}^{\infty}
\exp\!\Big(-c(\hbar \omega'-\Delta) -\frac{(\hbar \omega- \hbar \omega')^2}{2\sigma_E^2} \Big)\,d\omega', 
\qquad
c \equiv \frac{1}{2\alpha\sigma_Q^{2}}
\label{eq:J_app}
\end{equation}
where the Heavyside step function, makes the integration range $\hbar\omega' \in [\Delta,\infty]$. 
By completing the square, we can rewrite the exponent:
\begin{equation}
- c(\hbar\omega' - \Delta) -\frac{(\hbar\omega - \hbar\omega')^2}{2\sigma_E^2}
= -\frac{1}{2\sigma_E^2}(\hbar\omega')^2
+ \Big(\frac{\hbar\omega}{\sigma_E^2} - c\Big)\hbar\omega'
- \frac{(\hbar\omega)^2}{2\sigma_E^2} + c\Delta.
\end{equation}

We define $\mu \equiv \hbar\omega - c\,\sigma_E^2$, add and subtract $\mu^2$, such that the exponent becomes,
\begin{align}
- c(\hbar\omega' - \Delta) -\frac{(\hbar\omega - \hbar\omega')^2}{2\sigma_E^2}& = -\frac{1}{2\sigma_E}( (\hbar \omega')^2 -2\mu \hbar \omega' + \mu^2) + \frac{1}{2\sigma_E} (\mu^2 - (\hbar\omega)^2) + c\Delta, \\
   & = -\frac{(\hbar\omega' - \mu)^2}{2\sigma_E^2} - c\,\hbar\omega + \tfrac{1}{2}c^2\sigma_E^2 + c\Delta.
\end{align}

Hence,
\begin{equation}
    J =  \exp \left(-c\hbar\omega + \tfrac{1}{2}c^2\sigma_E^2 + c\Delta \right) 
\int_{\Delta}^{\infty}
\exp\!\Big[-\frac{(\hbar\omega' - \mu)^2}{2\sigma_E^2}\Big]
\,d\omega'.
\end{equation}

Using the standard Gaussian tail identity; 
\begin{equation}
    \int_{\Delta}^{\infty} \exp\!\Big[-\frac{(\hbar \omega'-\mu)^2}{2\sigma^2}\Big]d\omega' = \sigma_E\sqrt{\frac{\pi}{2}} \operatorname{erfc}\!\Big(\frac{\Delta-\mu}{\sqrt{2}\,\sigma_E}\Big),
\end{equation}
and we find
\begin{equation}
    J = e^{-c\hbar\omega + \tfrac{1}{2}c^2\sigma_E^2 + c\Delta}
\;\sigma_E\sqrt{\frac{\pi}{2}}\;
\operatorname{erfc}\!\Big(\frac{\Delta - \mu}{\sqrt{2}\,\sigma_E}\Big).
\end{equation}

Substituting back \(\mu = \hbar\omega - c\sigma_E^2\) gives the analytical expression, 
\begin{equation}
I(\textbf{Q}, \hbar\omega)
= A\frac{1}{4 \alpha \sigma_Q^2}\;
\exp\!\Big[\frac{1}{2\alpha \sigma_Q^2}(\Delta - \hbar\omega) + \tfrac{1}{2} \Big(\frac{ \sigma_E}{2 \alpha \sigma_Q^2} \Big)^2\Big]\;
\operatorname{erfc}\!\left(\frac{\Delta - \hbar\omega + \Big(\frac{ \sigma_E}{2 \alpha \sigma_Q^2} \Big)^2}{\sqrt{2}\,\sigma_E}\right) +B
\label{eq:analytical_app}
\end{equation}
This form is valid for all $\hbar\omega$. The Heaviside function in the base, $f(\omega)$ in eq. \eqref{eq:heavyside_app}, ensures support only for $\hbar\omega \ge \Delta$, but the convolution itself yields a smooth non-zero tail for all $\hbar\omega$ given by the erfc-function.
Comparing the numerical expression in eq. \eqref{eq:fitting_func_app} and the analytical function in eq. \eqref{eq:analytical_app} gives an agreement better than 0.01\% , the tiny difference we ascribe to the finite grid resolution in the numerical integral. 

\section{Limitations of the convoluted gap function}\label{app:limitations}
Our convoluted-gap model has some limitations of when the expression is valid. The assumptions used have to be fulfilled to get the best possible fit.

\subsection{Accuracy and Limits of the Parabolic Approximation}\label{app:limitations_parab}
\begin{figure}[ht!]
    \centering
    \includegraphics[width=0.5\linewidth]{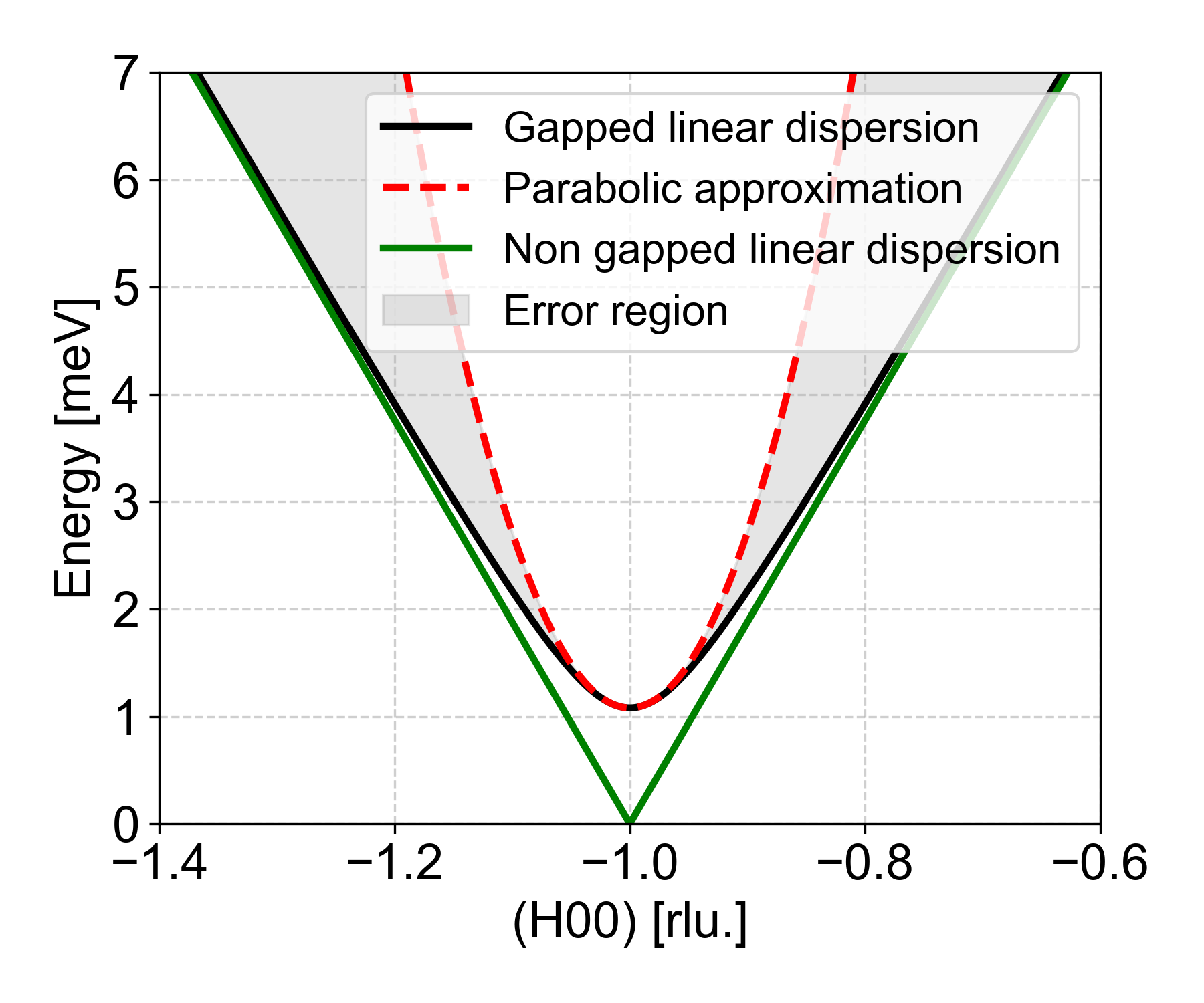}
    \caption{The gapped linear dispersion, eq.~\eqref{eq:AFM_dispersion} (black line) is approximated by a parabola, eq.~\eqref{eq:parabola} (dashed red line). The shaded area indicates the error region of this approximation. In green, we plot the non-gapped linear dispersion, to show the similarity to the square-root behaviour.}
    \label{fig:error_region}
\end{figure}

Our approximated convoluted gap function in eq. \eqref{eq:analytical} will overestimate the tail above the gap position due to the approximation of the linear excitation in eq. \eqref{eq:AFM_dispersion} following the shape of the parabola in eq. \eqref{eq:parabola}. In figure \ref{fig:error_region}, we illustrate the error region of the approximation, as well as the non-gapped linear dispersion. 

Physically, the parabolic approximation is reliable for describing the near-gap region, where the dispersion has a constant curvature, but not for the outer regions where the dispersion becomes more linear due to the square-root dependence. In those tails, the approximation exaggerates the dispersion value and leads to an overprediction of the total signal when integrated. The degree to which this overestimation is apparent depends on the instrumental resolution. 

In the case of a focused TAS, the instrumental resolution is typically anisotropic, with two wide and one narrow component in momentum transfer. This geometry produces a resolution ellipsoid that is elongated along certain reciprocal-space directions, effectively averaging the scattering intensity over a broad region in $Q$ while maintaining sharp definition along one axis. As a result, the measured dispersion appears smoother and more parabolic, since the averaging suppresses the intrinsic square-root flattening of the linear dispersion. The broad $Q$ acceptance leads to a systematic overestimation of the intensity in the high-energy tails, while the narrow component preserves accurate determination of the gap position. Thus, in focused TAS measurements, the parabolic approximation often provides an excellent apparent fit near the gap.

\subsection{Isotropic spin wave velocity}
We assume an isotropic spin wave velocity, $a$. If the real material has anisotropic spin-wave velocities $a_i$ along different crystal axes, then
\begin{equation}
    E(\textbf{Q}) = \sqrt{\sum_i a_i(\textbf{Q}_i-\textbf{Q}_0)^2 + \Delta^2}.
\end{equation}
The small-Q expansion then yields an anisotropic quadratic form $E \approx \Delta + \frac{1}{2\Delta}\sum_i a_i^2(\textbf{Q}_i-\textbf{Q}_0)^2$, i.e. elliptical constant-energy contours instead of circular. Thus, the isotropic parabola is only valid when the anisotropy is small.

\subsection{The Bose factor}
Thermal fluctuations can affect the spin wave spectrum, which is described in the thermal population factor (the Bose factor), $n(\omega)+1 = 1/(1-\exp(-\hbar\omega/(k_B T))$, where $k_B$ is the Boltzmann constant and $T$ is the temperature. 
We assume a dispersion spectrum at low temperatures, where thermal fluctuations does not affect the spectrum. The Bose factor at low temperatures is $k_B T \ll \hbar \omega$, where $n(\omega)+1\approx1$ and we get eq. \eqref{eq:analytical}. 

In the high-temperature (classical) limit, the Bose factor becomes approximately $\hbar\omega/(k_B T)$. This factor reflects the occupation of excited states rather than the underlying spin dynamics. Thus, at high-temperatures the factor $(k_B T)/\hbar\omega$ is added to the integral in eq. \eqref{eq:fitting_func}, which can not be written in an analytic form. 
However, instead of having $J$ in eq. \eqref{eq:J_app}, we can define $K$ as; 
\begin{equation}
    K(\mu) = \int_{\Delta}^{\infty}  \frac{k_B T}{\hbar \omega'}
\exp\!\Big( -\frac{(\hbar \omega'- \mu)^2}{2\sigma_E^2} \Big)\,d\omega', 
\qquad
\mu = \hbar \omega - c\sigma_E.
\end{equation}

\section{McStas Model}\label{app:McStas_model}
For the dispersion in MnF$_2$, we have implemented an analytical model from Ref.~\cite{Yamani_2010} in McStas, see Ref.~\cite{Schack2025}. We simulate data from MnF$_2$ in McStas. For simulating the spectrometer, we use a doubly-focusing cold-neutron TAS model with
\begin{itemize}
  \item A $15 \times 15$~cm$^2$ source.
  \item A $2\times 2$~cm$^2$ slit before the monochromator as a virtual source.
  \item A curved $20\times 15$~cm$^2$ PG monochromator with a horizontal curvature of 2.7~m and a vertical curvature of 0.9~m. This is positioned 1.5~m before the sample. 
  \item A slit with a variable width to control the resolution; 0.9~m before the sample.
  \item A cylindrical sample, 1.25~cm tall, 1~cm diameter, representing AFM spin waves from MnF$_2$. We here use the new \verb|SpinWave_BCO| component in McStas \cite{Schack2025}.
  \item A second slit with the same variable width; 0.9~m after the sample.
  \item A $15\times 15$~cm$^2$ PG analyzer with a horizontal curvature of 1.67~m and no vertical curvature. The analyzer is positioned 1~m from the sample and is set to reflect at a constant $E_{\rm f} = 5.0$~meV. 
  \item A single energy-sensitive detector, $2.5 \times 1.25$~cm$^2$.
\end{itemize}

In the simulations, the source energy range is limited to avoid higher order scattering from the monochromator. Likewise, the detected energy range is limited to avoid higher order scattering from the analyzer.

\begin{figure}[ht!]
    \centering
    \includegraphics[width=.7\linewidth]{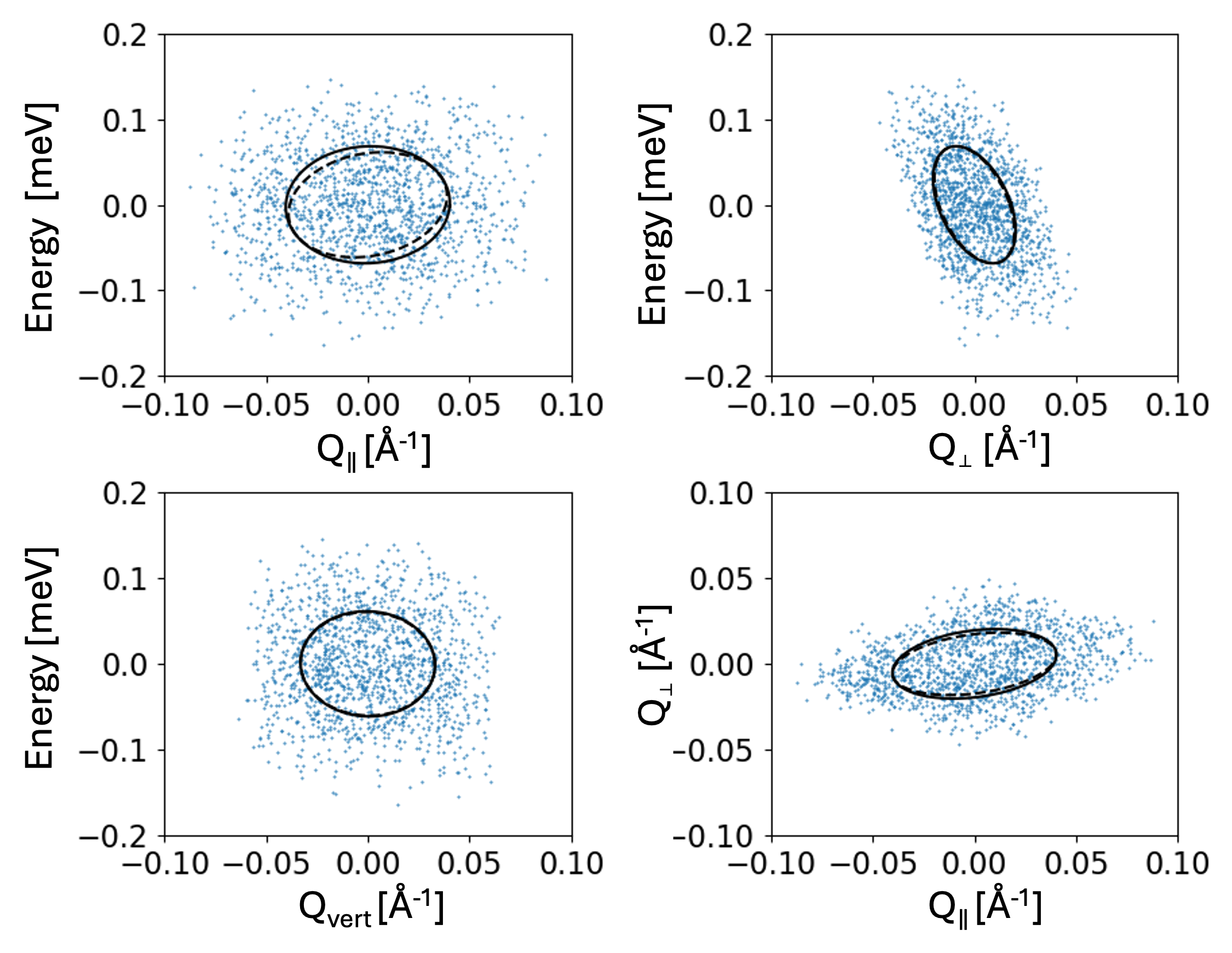}
    \caption{Simulated resolution ellipsoid of the McStas TAS model with a slit width of 5~cm. The blue dots are scattered neutrons and the solid black ellipsis are the projected resolution along the energy axis, while the dashed indicate the resolution at zero energy transfer.}
    \label{fig:res_elipsis_TAS40}
\end{figure}

We simulate the resolution function of the instrument by replacing the sample and detector with the McStas components \verb|Res_sample| and \verb|Res_Monitor|, respectively. The method is described in Ref.~\cite{Lefmann_2000}. The instrumental resolution at nine different slit widths is reported in Table \ref{tab:resolution_McStas}. As an example, the resolution ellipses for the setting with slit widths of 5~cm is presented in figure~\ref{fig:res_elipsis_TAS40}, showing 4 different projected resolutions.

\section{Calculating the resolutions from the covariance matrix}\label{app:covariance_matrix}
The instrumental resolution of a neutron spectrometer can be fully described by a covariance matrix $\mathbf{C}$, which characterizes the uncertainties and correlations between the four measured quantities: the three components of the momentum transfer $\mathbf{Q}$ and the energy transfer $E$.

In the laboratory (Cartesian) coordinate system, the resolution covariance matrix is written as
\begin{equation}
\mathbf{C} =
\begin{pmatrix}
\langle \Delta Q_x^2 \rangle &
\langle \Delta Q_x \Delta Q_y \rangle &
\langle \Delta Q_x \Delta Q_z \rangle &
\langle \Delta Q_x \Delta E \rangle \\[6pt]
\langle \Delta Q_y \Delta Q_x \rangle &
\langle \Delta Q_y^2 \rangle &
\langle \Delta Q_y \Delta Q_z \rangle &
\langle \Delta Q_y \Delta E \rangle \\[6pt]
\langle \Delta Q_z \Delta Q_x \rangle &
\langle \Delta Q_z \Delta Q_y \rangle &
\langle \Delta Q_z^2 \rangle &
\langle \Delta Q_z \Delta E \rangle \\[6pt]
\langle \Delta E \Delta Q_x \rangle &
\langle \Delta E \Delta Q_y \rangle &
\langle \Delta E \Delta Q_z \rangle &
\langle \Delta E^2 \rangle
\end{pmatrix}.
\label{eq:covariance}
\end{equation}

Each element $\langle \Delta A \Delta B \rangle$ represents the covariance between parameters $A$ and $B$. Each diagonal term gives the variance (square of $1\sigma$ uncertainty) of $Q_x$, $Q_y$, $Q_z$, and $E$, while the off-diagonal terms quantify how uncertainties in these quantities are correlated due to the geometry and kinematics of the instrument.

\paragraph{Coordinate transformations.}
Although $\mathbf{C}$ is often expressed in the laboratory coordinate system $(Q_x, Q_y, Q_z, E)$, it is frequently more useful to represent it in a local coordinate frame defined relative to the measured scattering geometry, such as $(Q_{\parallel}, Q_{\perp}, Q_{\mathrm{vert}}, E)$.
Here:
\begin{itemize}
    \item $Q_{\parallel}$ is parallel to the nominal \textbf{Q}-direction. 
  \item $Q_{\perp}$ lies in the scattering plane, but perpendicular to $Q_{\parallel}$. 
  \item $Q_{\mathrm{vert}}$ is perpendicular to the scattering plane.
\end{itemize}

The transformation from the laboratory frame to the local frame is given by
\begin{equation}
    \mathbf{C}' = \mathbf{R}^\mathrm{T} \, \mathbf{C} \, \mathbf{R},
\end{equation}
where $\mathbf{R}$ is a rotation matrix whose columns define the local basis vectors expressed in laboratory coordinates. This transformation rotates the covariance matrix into a frame that is physically aligned with the measurement direction.

Expressing the covariance matrix in $(Q_{\parallel}, Q_{\perp}, Q_{\mathrm{vert}}, E)$ provides a more intuitive picture of the instrumental resolution in reciprocal space and energy. In this frame, the principal axes of the resolution ellipsoid typically align more closely with physically meaningful directions—such as along the dispersion relation or perpendicular to the scattering plane—making it easier to visualize how instrumental uncertainties contribute to the measured signal.

\subsection{Determination of the principal axes of the \texorpdfstring{$Q$}{Q}-resolution in the scattering plane}
To visualize the resolution in the scattering plane, e.g. $(Q_{x}, Q_{y})$, the full
covariance matrix is projected using a projection matrix $\mathbf{P}$, defined such that
\begin{equation}
    \mathbf{C}_{\mathrm{proj}} = \mathbf{P}^\mathrm{T} \, \mathbf{C} \, \mathbf{P},
\end{equation}
where $\mathbf{P}$ selects the desired subspace (e.g. $(Q_{x}, Q_{y})$-plane).

The principal axes of the resolution ellipse in this plane are obtained by diagonalizing the projected covariance matrix:
\begin{equation}
    \mathbf{C}_{\mathrm{proj}} \, \mathbf{v}_i = \lambda_i \, \mathbf{v}_i,
\end{equation}
where $\lambda_i$ and $\mathbf{v}_i$ are the eigenvalues and eigenvectors, respectively. Each eigenvector $\mathbf{v}_i$ defines the orientation of a principal axis of the ellipse, while the corresponding eigenvalue
$\lambda_i$ gives the variance along that direction.

The $1\sigma$ widths along the principal axes are
\begin{equation}
    \sigma_i = \sqrt{\lambda_i},
\end{equation}
and the full width at half maximum (FWHM) values are
\begin{equation}
    \mathrm{FWHM}_i = 2 \sqrt{2 \ln 2} \, \sigma_i.
\end{equation}
These are the values we report as the resolution of the instrument.

\subsection{Energy Resolution from the Covariance Matrix}
The instrumental covariance matrix in eq.~\eqref{eq:covariance} can be rewritten, such that the momentum transfer is written as $\mathbf{Q} = (Q_x, Q_y, Q_z)$:
\begin{equation}
\mathbf{M} =
\begin{pmatrix}
\mathbf{M}_{QQ} & \mathbf{M}_{QE} \\
\mathbf{M}_{EQ} & M_{EE}
\end{pmatrix},
\end{equation}
where $\mathbf{M}_{QQ}$ is a $3\times3$ covariance submatrix in momentum space, $\mathbf{M}_{QE} = \mathbf{M}_{EQ}^\mathrm{T}$ contains the cross-covariances between $\mathbf{Q}$ and $E$, and $M_{EE}$ is the variance of the energy transfer.

Two different definitions of the energy resolution can be derived from the covariance matrix, $\mathbf{M}$, depending on whether the momentum $\mathbf{Q}$ is considered to be fixed or not. Whether one needs to use one or the other, depends on the steepness of the dispersion and the size of the $Q$-resolution, see figure \ref{fig:resolution_assumptions} right. The intrinsic energy resolution at a specific Q (orange in figure), $\sigma_{E|\mathbf{Q}}$, is typically smaller than the projected resolution (green in figure), $\sigma_E$, because fixing $\mathbf{Q}$ removes the influence of the $\mathbf{Q}$--$E$ correlations that tilt the resolution ellipsoid in $(\mathbf{Q},E)$-space.

\paragraph{(1) Intrinsic energy resolution at specific Q.}
If the momentum transfer $\mathbf{Q}$ is fixed, the intrinsic energy-resolution variance is obtained from the conditional variance of $E$ given $\mathbf{Q}$:
\begin{equation}
    \mathrm{Var}(E \mid \mathbf{Q})
  = M_{EE} - \mathbf{M}_{EQ}\,\mathbf{M}_{QQ}^{-1}\,\mathbf{M}_{QE}.
\end{equation}

The corresponding one-standard-deviation width and the full width at half maximum (FWHM) is
\begin{equation}
    \sigma_{E|\mathbf{Q}} = \sqrt{\mathrm{Var}(E \mid \mathbf{Q})}, \qquad
    \mathrm{FWHM}_{E|\mathbf{Q}} = 2\sqrt{2\ln2}\,\sigma_{E|\mathbf{Q}}.
\end{equation}
This represents the intrinsic energy broadening of the instrument, i.e.\ the spread in $E$ when we are at a specific $\mathbf{Q}$, like at the gap.

\paragraph{(2) Projected energy resolution.}
If the momentum transfer is not fixed, the total projected energy-resolution variance is simply the marginal variance of $E$:
\begin{equation}
    \mathrm{Var}(E) = M_{EE},
\end{equation}
with
\begin{equation}
    \sigma_E = \sqrt{M_{EE}}, \qquad
    \mathrm{FWHM}_E = 2\sqrt{2\ln2}\,\sigma_E.
\end{equation}
This quantity includes both the intrinsic energy broadening and the additional contribution from the correlations between $\mathbf{Q}$ and $E$. It therefore represents the overall apparent energy resolution observed when the measurement integrates over the finite $\mathbf{Q}$ range the $Q$-resolution includes.

\section{Fitting the simulated data}\label{app:fitting_param}
\begin{figure}[ht!]
    \centering
    \includegraphics[width=0.5\linewidth]{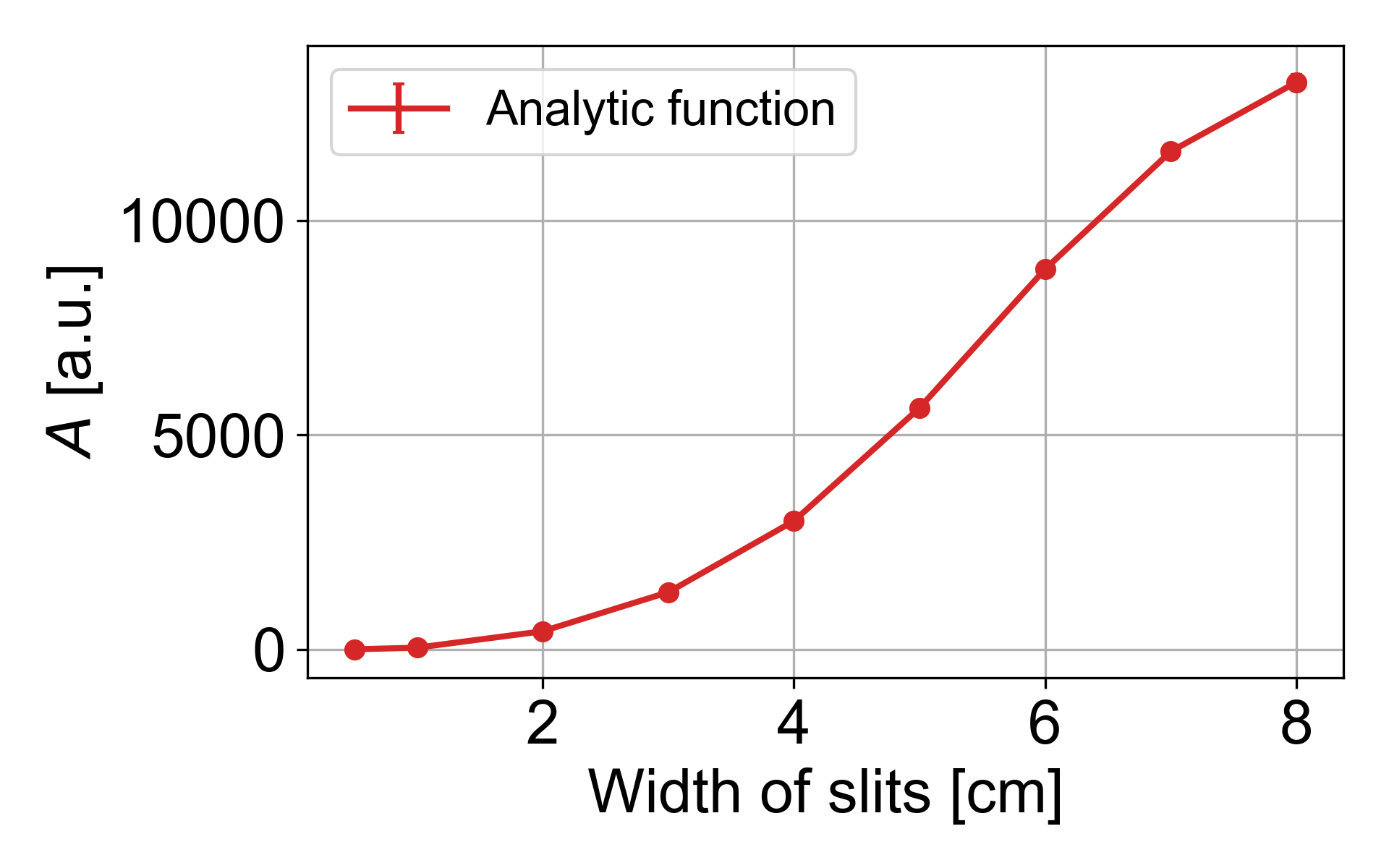}
    \caption{Normalisation constant $A$ fitted to the McStas data to the analytical fitting function in eq. \eqref{eq:analytical}.}
    \label{fig:A}
\end{figure}

\begin{figure}[H]
    \centering

    \begin{subfigure}{0.47\textwidth}
        \includegraphics[width=\textwidth]{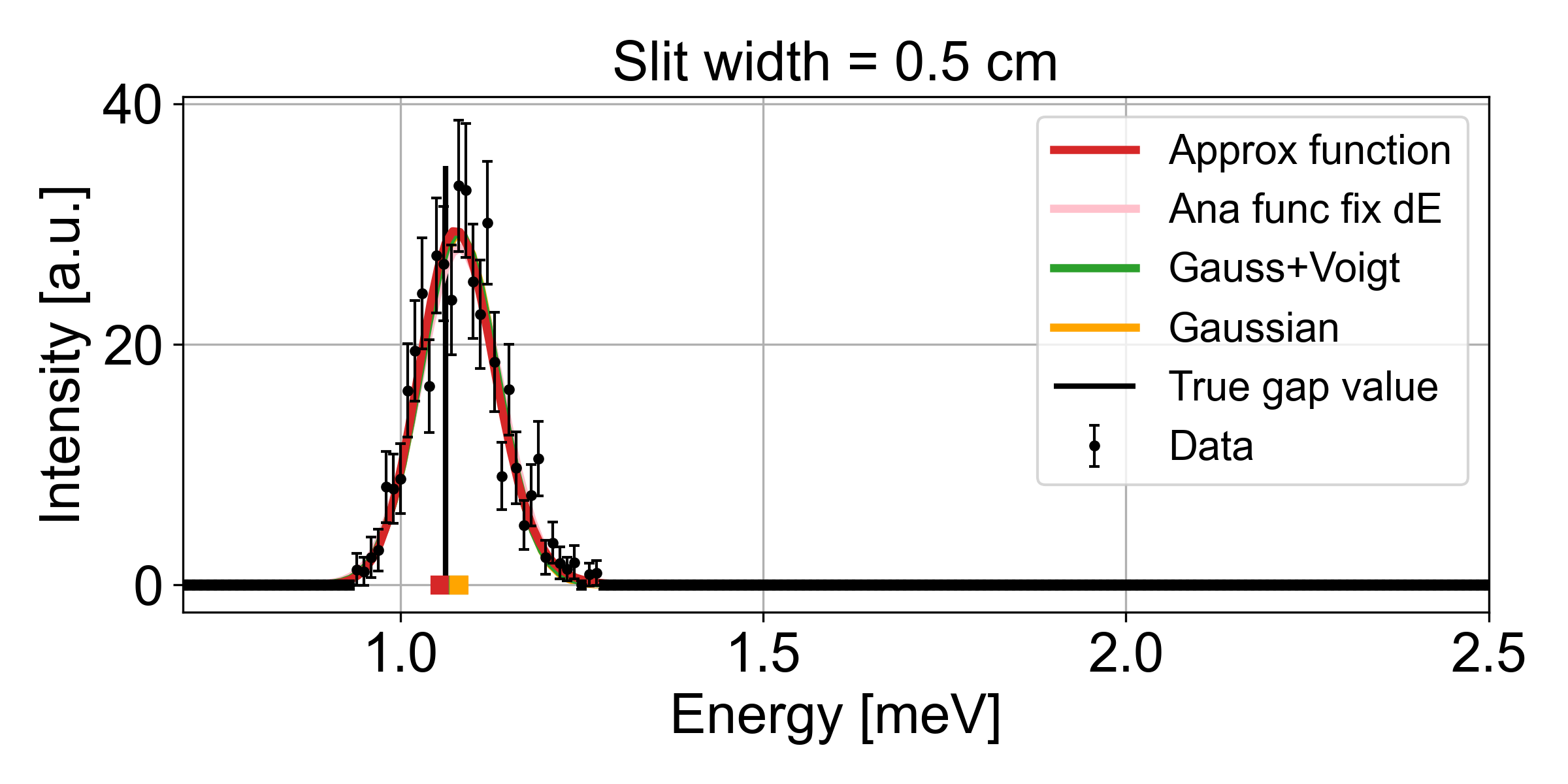}
    \end{subfigure}
    \hfill
    \begin{subfigure}{0.47\textwidth}
        \includegraphics[width=\textwidth]{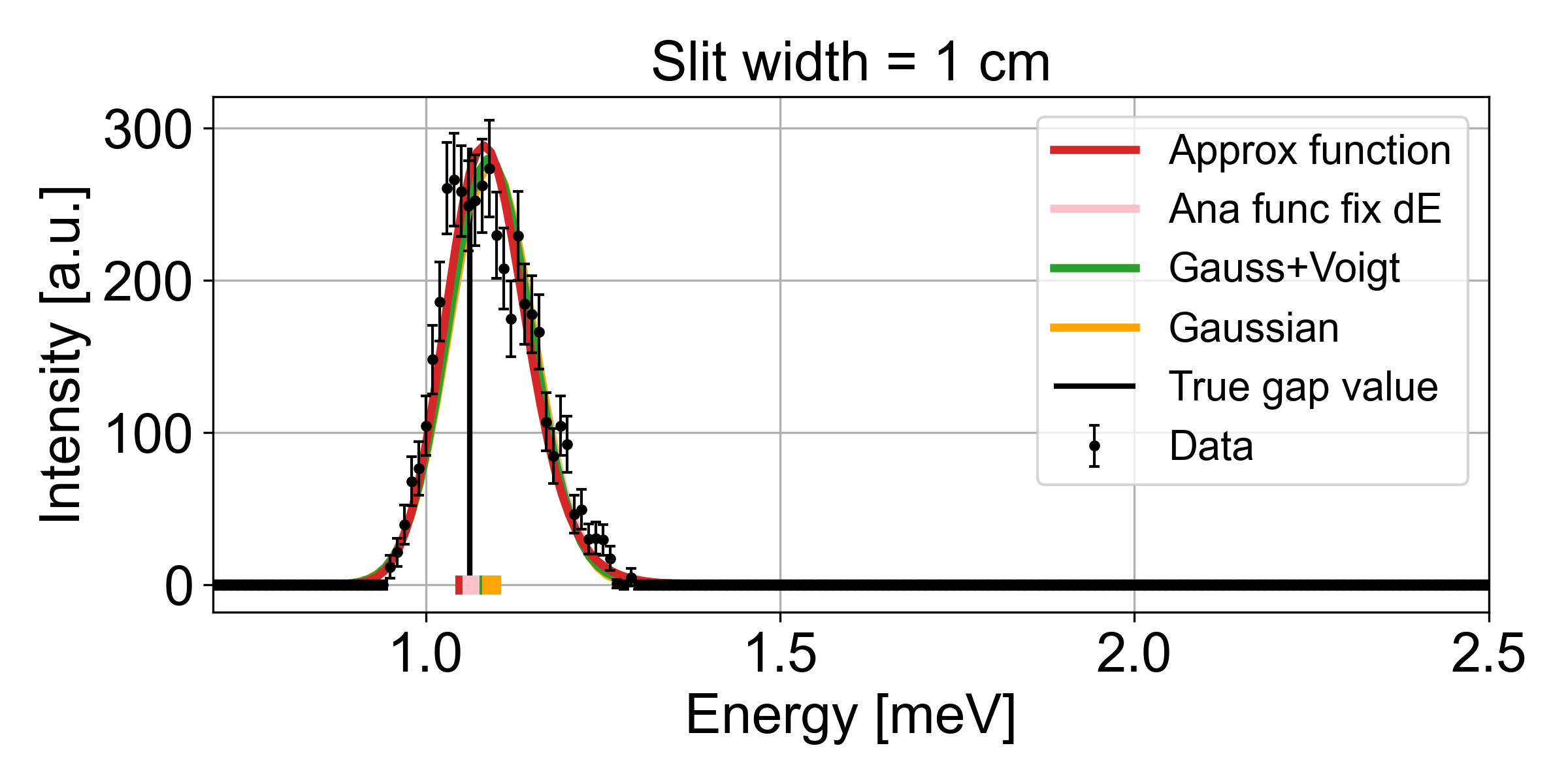}
    \end{subfigure}

    \vskip 0.4em
    \begin{subfigure}{0.47\textwidth}
        \includegraphics[width=\textwidth]{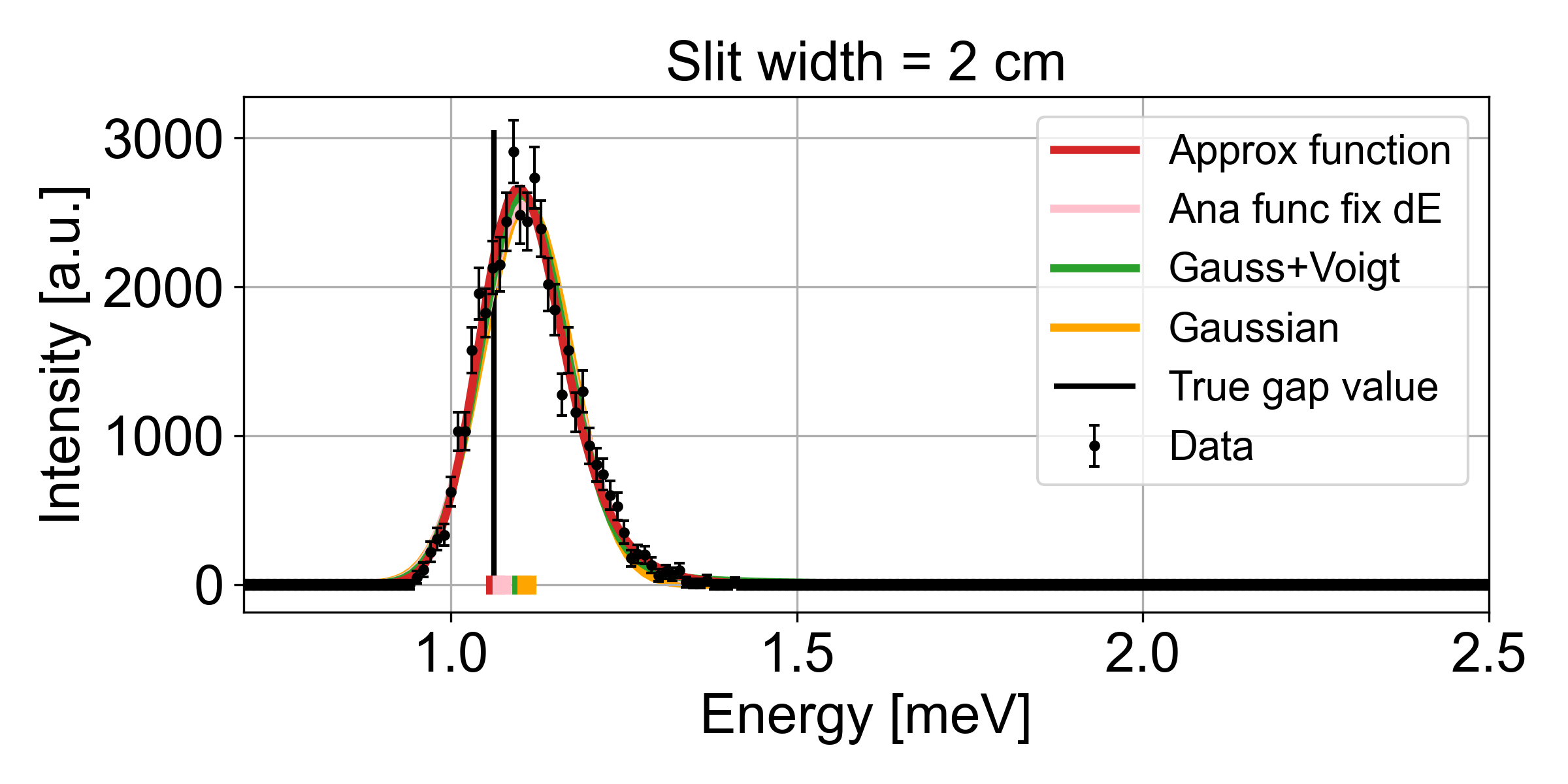}
    \end{subfigure}
    \hfill
    \begin{subfigure}{0.47\textwidth}
        \includegraphics[width=\textwidth]{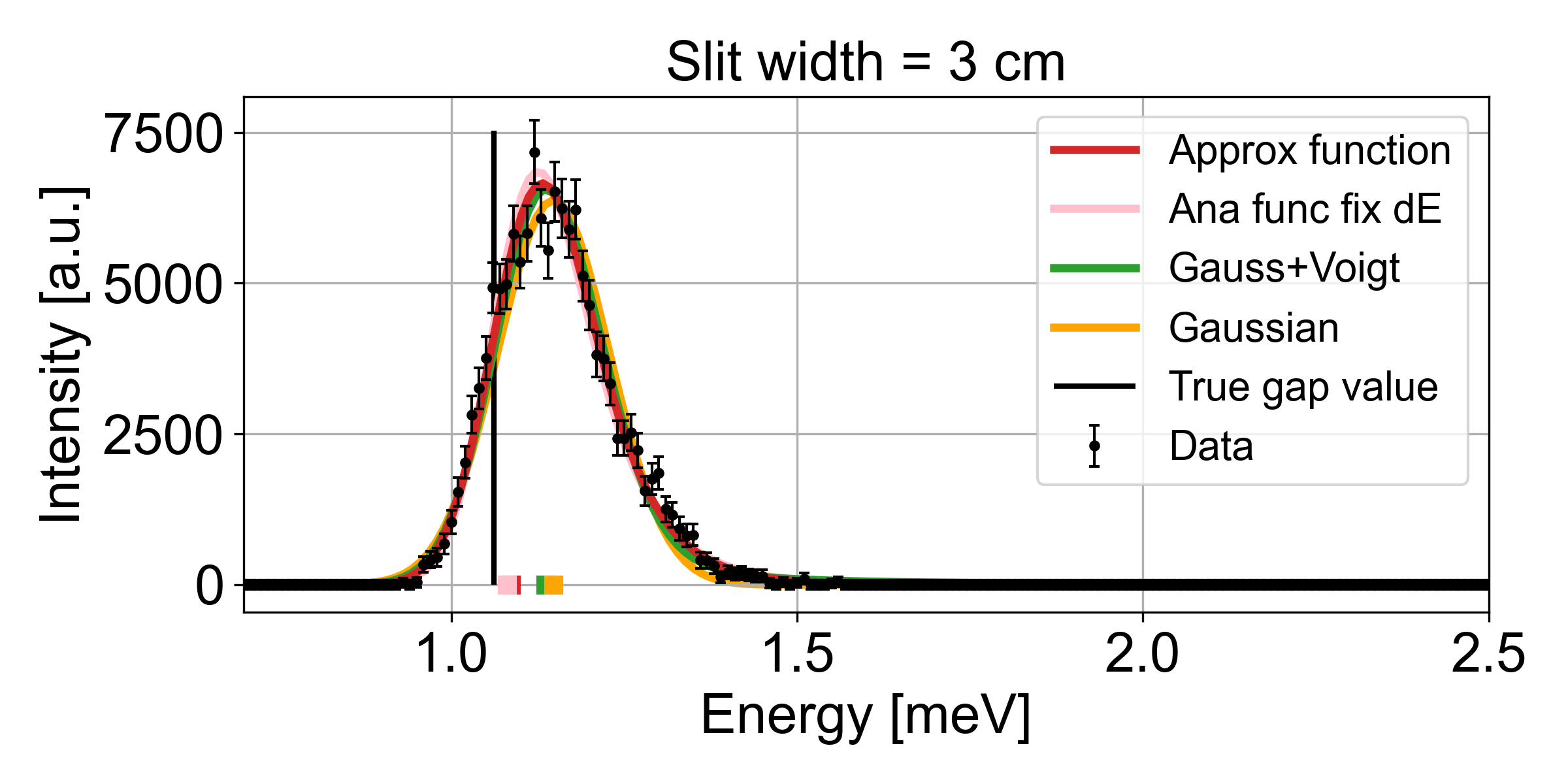}
    \end{subfigure}

    \vskip 0.4em
    \begin{subfigure}{0.47\textwidth}
        \includegraphics[width=\textwidth]{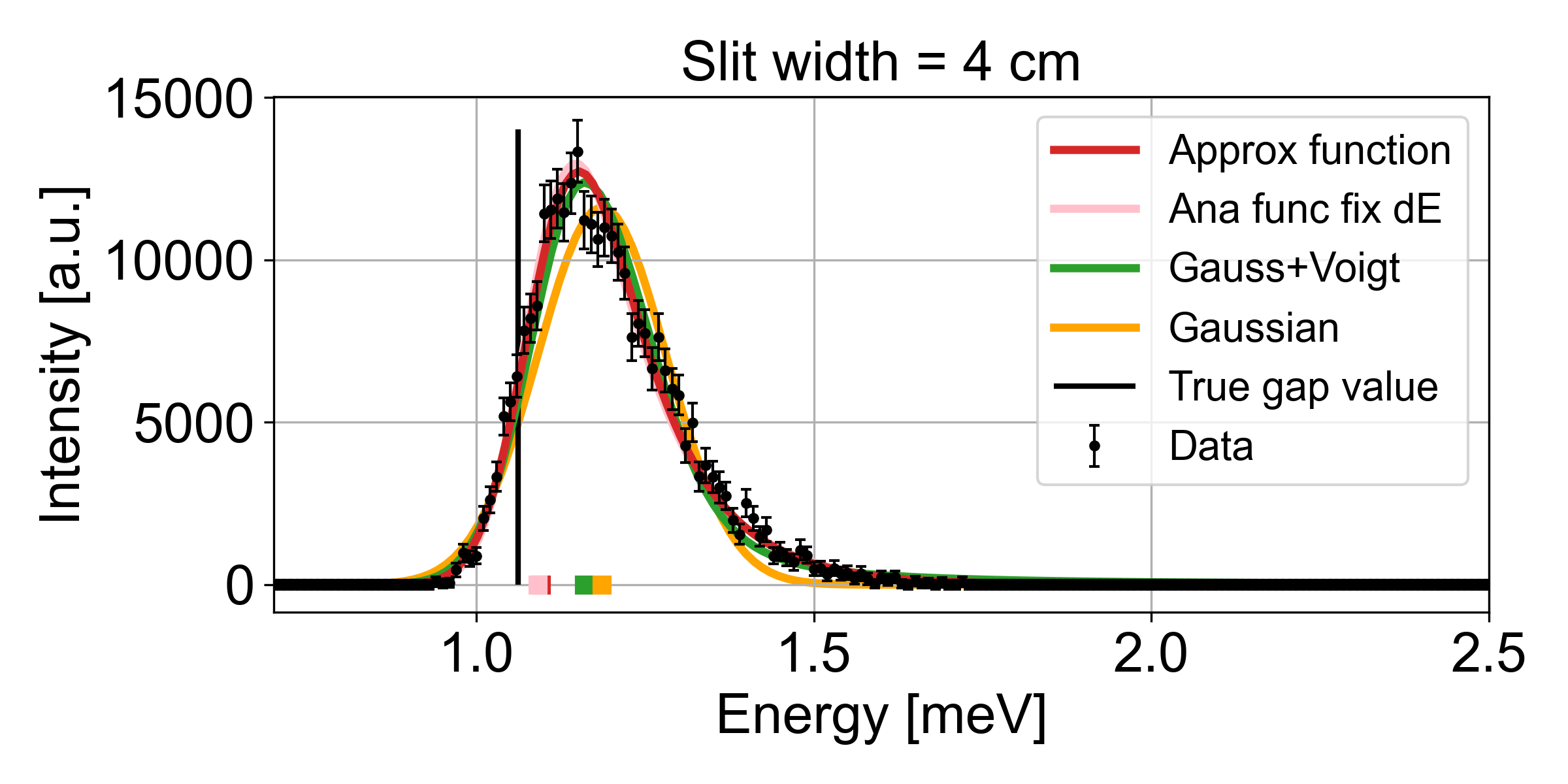}
    \end{subfigure}
    \hfill
    \begin{subfigure}[b]{0.47\textwidth}
        \includegraphics[width=\textwidth]{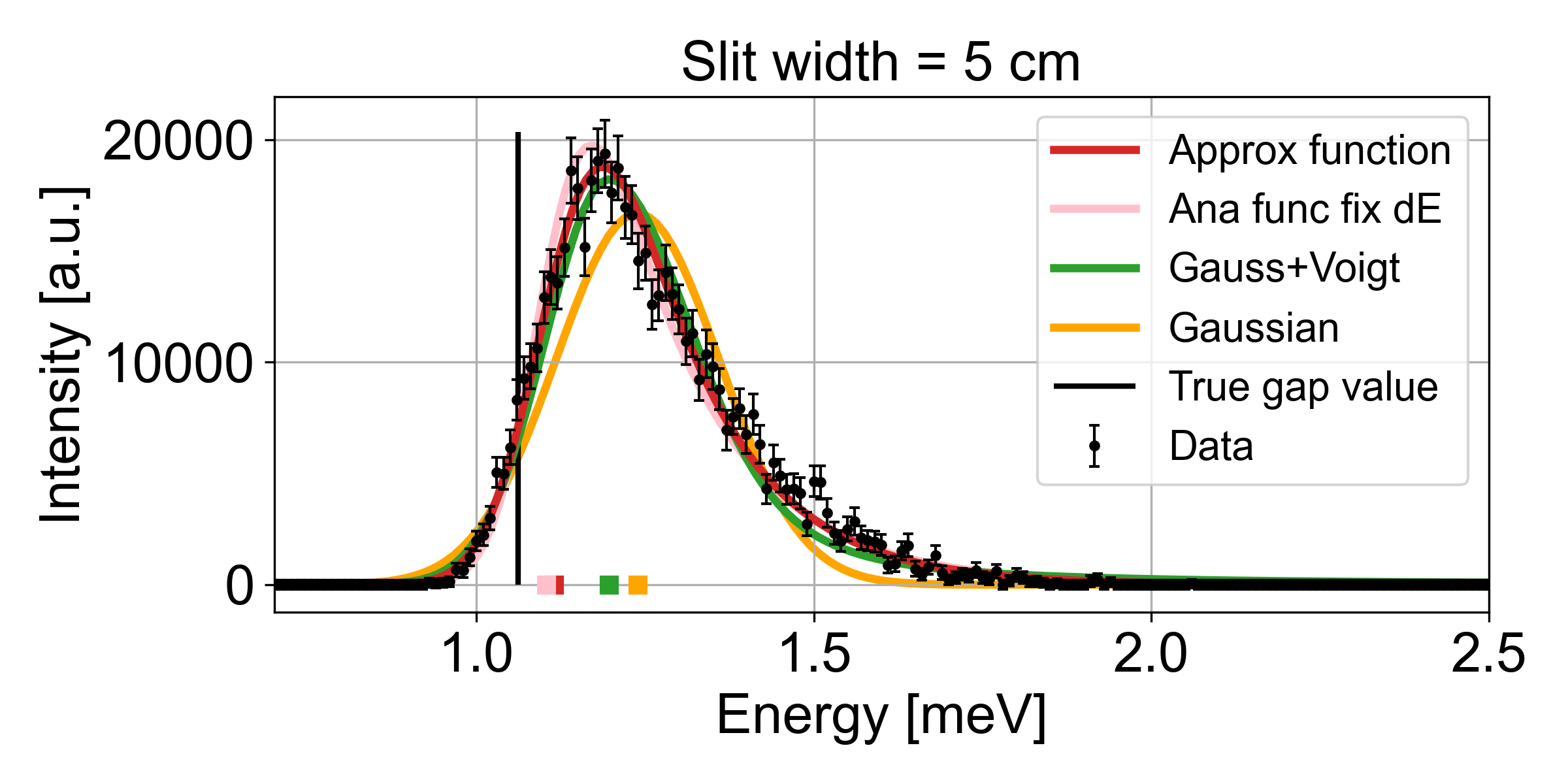}
    \end{subfigure}

    \vskip 0.4em

    \begin{subfigure}{0.47\textwidth}
        \includegraphics[width=\textwidth]{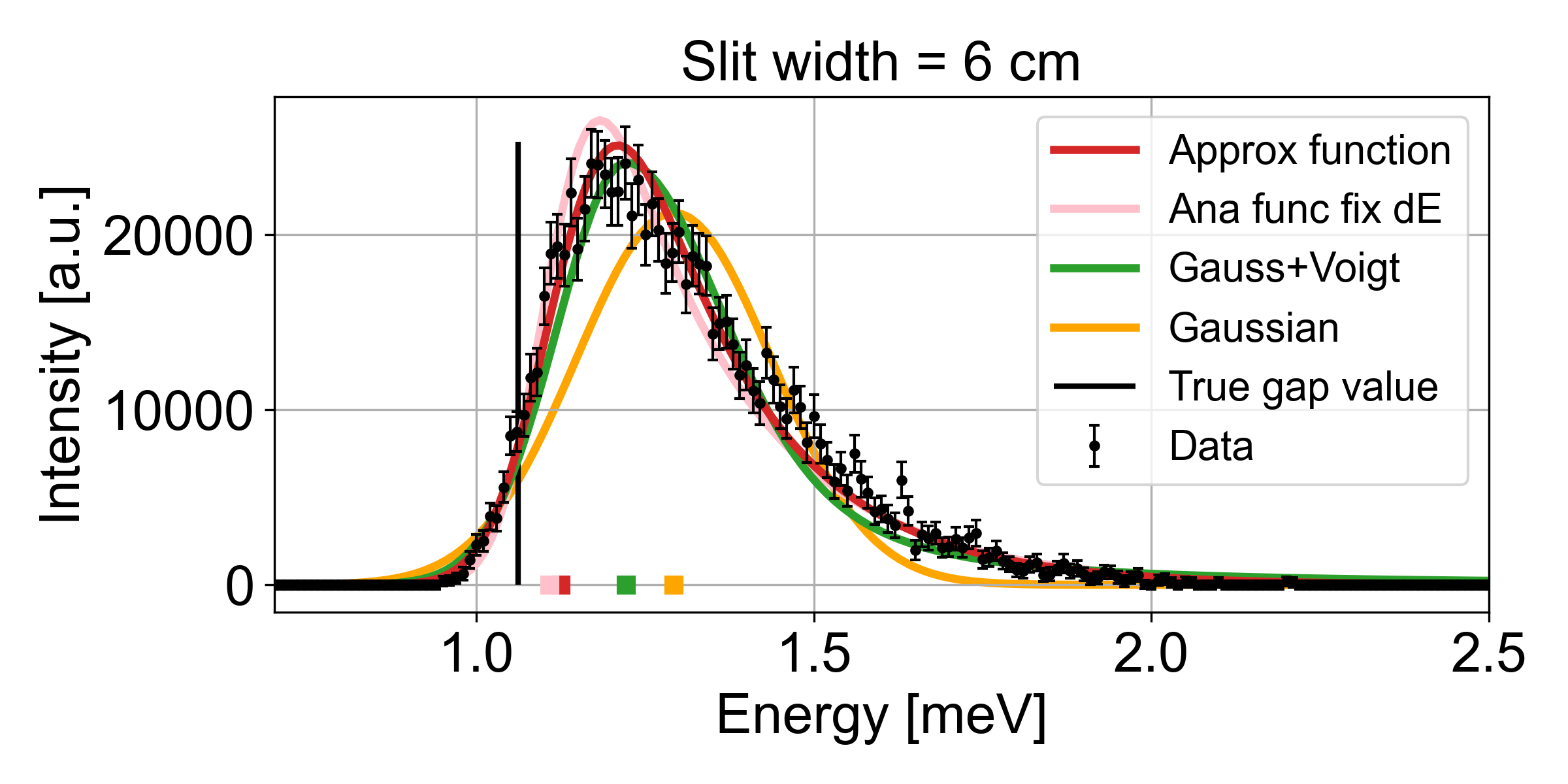}
    \end{subfigure}
    \hfill
    \begin{subfigure}{0.47\textwidth}
        \includegraphics[width=\textwidth]{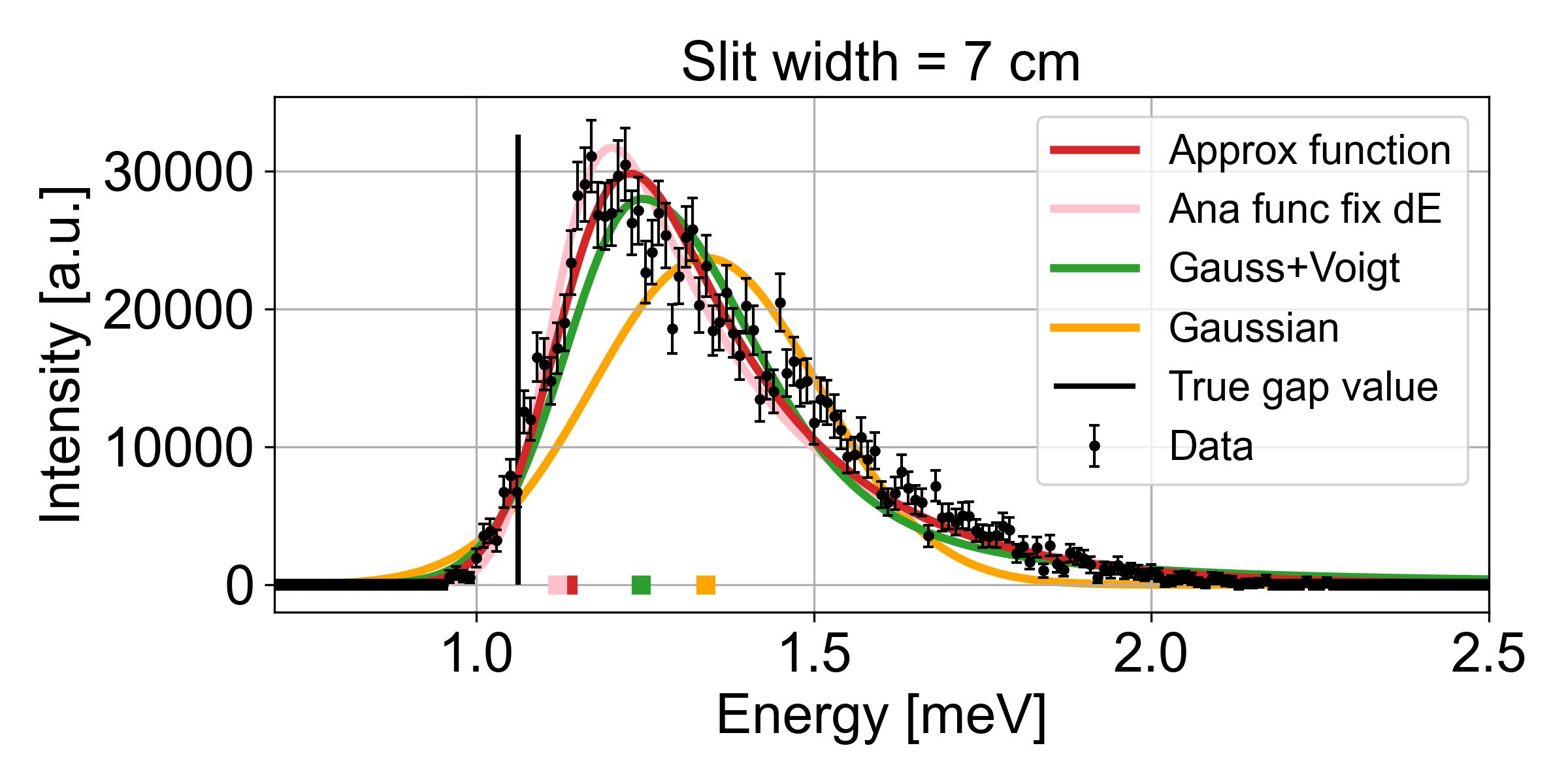}
    \end{subfigure}

    \vskip 0.4em

    \begin{subfigure}{0.47\textwidth}
        \includegraphics[width=\textwidth]{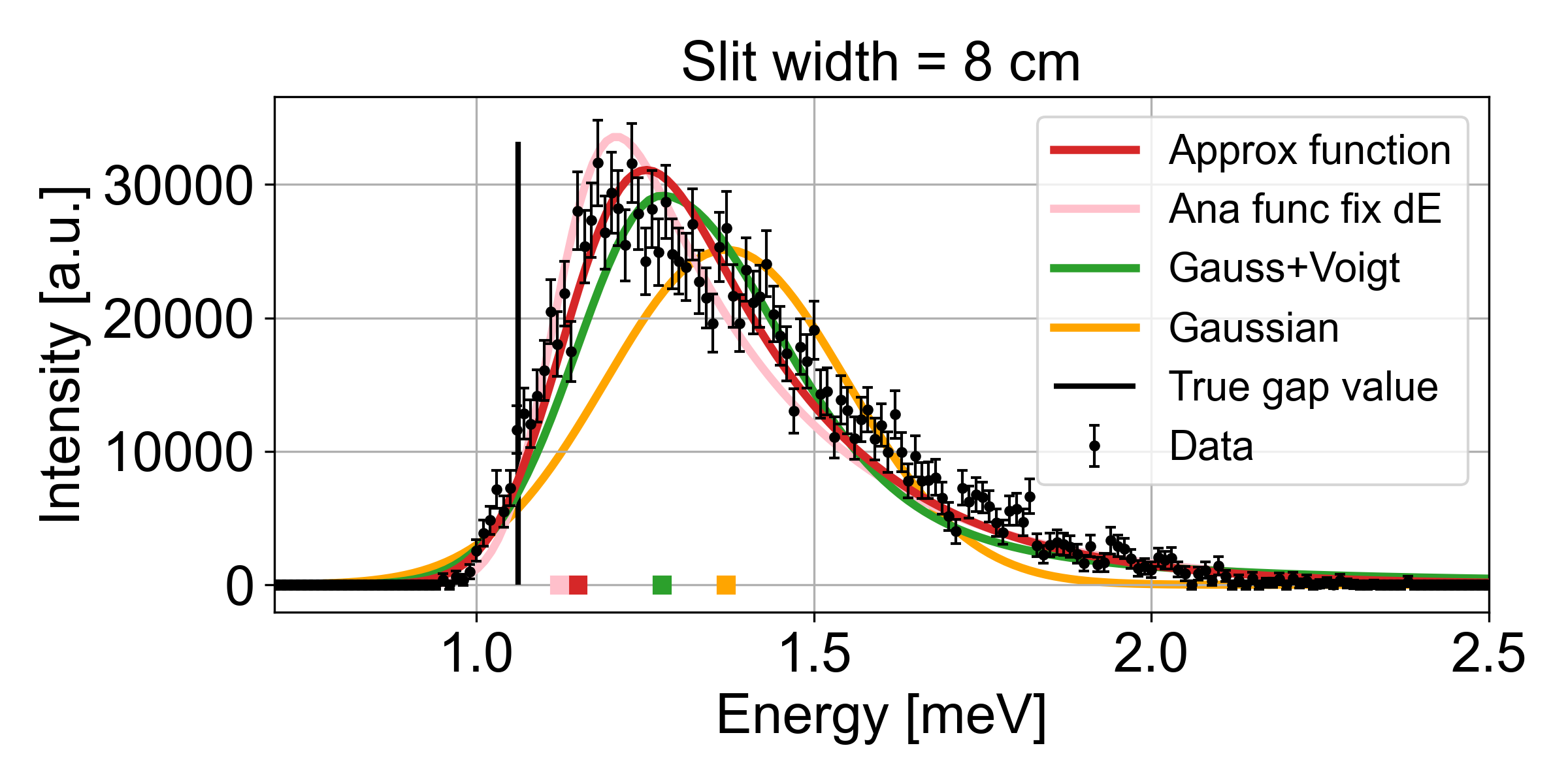}
    \end{subfigure}

    \caption{Overview of all simulated constant-Q cuts at the gap varying slit width from 0.5-8 cm, which effectively varies the $Q$-resolution. The black errorbars are the simulated data and have been fitted with a Gaussian (yellow line), a combined Gaussian plus Voigt in eq.~\ref{eq:Gauss_Voigt} (green line) and our approximate convoluted-gap function eq.~\ref{eq:analytical} (red line). We have also fitted with a fixed value of the energy resolution, $\sigma_E$, shown in pink. The coloured squares at zero indicates the fitted gap positions.}
    \label{fig:all_Mcstas_fits}
\end{figure}

As mentioned in the main text, we have simulated constant-Q cuts at the gap at varying slit width, which effectively varies the $Q$-resolution. All simulated results have been fitted with our analytic function \eqref{eq:analytical}, a Gaussian lineshape and a combined Gaussian plus Voigt function (see appendix \ref{app:Gauss+Voigt}). All fits can be seen in figure \ref{fig:all_Mcstas_fits}, these are not normalised. The fitted parameters are plotted in the main text; the gap size (figure \ref{fig:McStats_fits}b), the $Q$-resolution and energy resolution (figure \ref{fig:McStats_fit_resolution}) and the $\chi^2$ value (figure \ref{fig:McStats_fits}c). Additionally, we also fit the normalisation constant $A$, which can be seen in figure \ref{fig:A}. Noticing the value of $A$ as a function of slit width, it is clear the larger the slit width, the more neutrons in the detector, up to a saturation value approx.~8~cm slit width.

\section{Fixed energy resolution in fitting the CAMEA data}\label{app:CAMEA_fit_fix_dE}
\begin{figure}[ht!]
    \centering
    \includegraphics[width=0.5\linewidth]{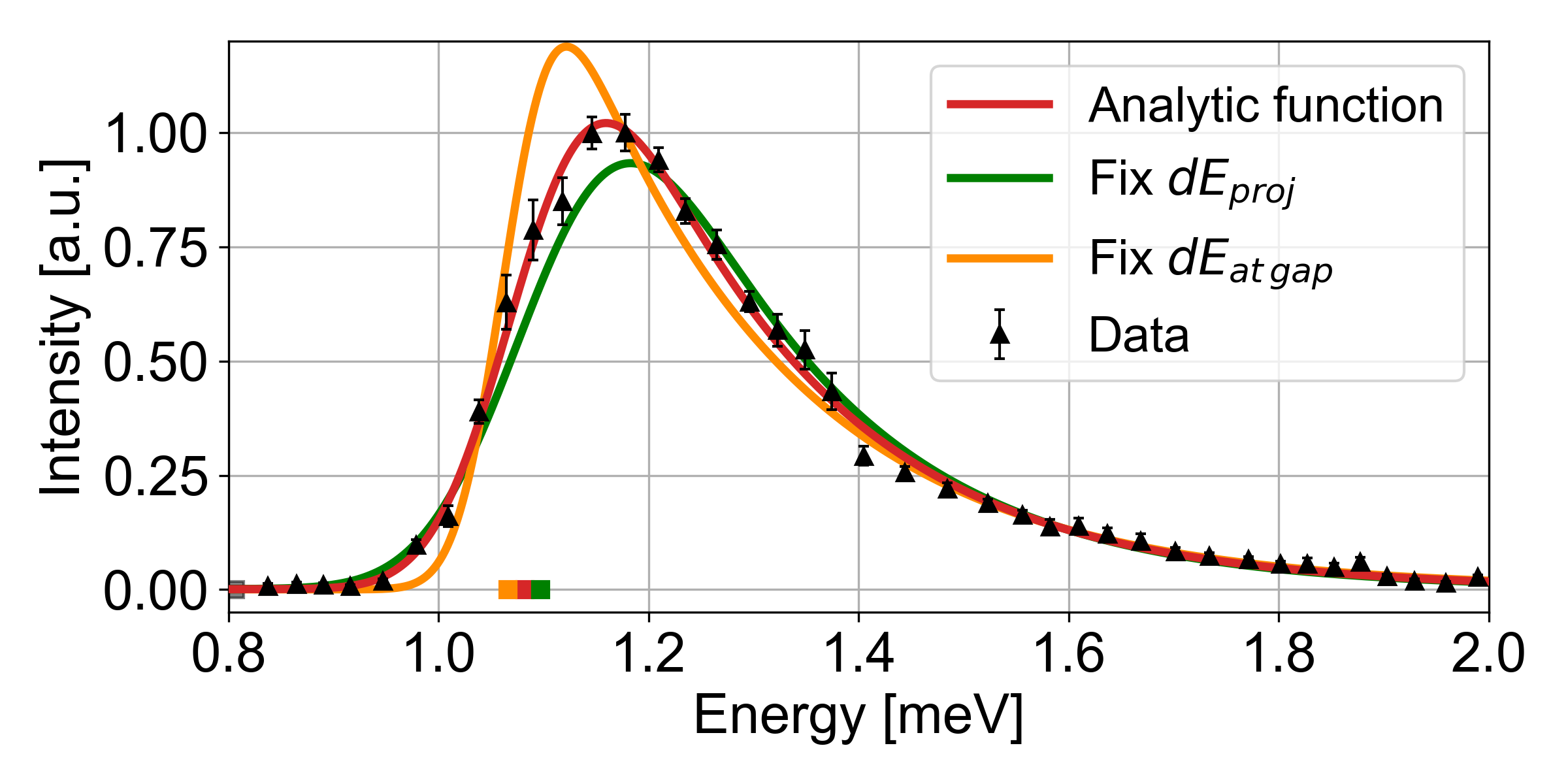}
    \caption{Testing the energy resolution found from the CAMEA data. In red, we fit the data (black errorbars) with our approximate convoluted-gap function eq. \eqref{eq:analytical} keeping $\sigma_E$ free. In green, we fix $\sigma_E$ to the projected energy resolution, while in orange, we fix $\sigma_E$ to the intrinsic energy resolution at the gap.}
    \label{fig:CAMEA_fix_energy_res}
\end{figure}

We fit our approximate convoluted-gap function, eq.~\eqref{eq:analytical}, to the experimental CAMEA data in figure \ref{fig:CAMEA_fix_energy_res} with the spin wave velocity, $a$, as the only fixed parameter. From the fit, we get a gap size of $\Delta=1.082(2)$~meV, $Q$-resolution $\sigma_Q=0.0308(2)$~Å$^{-1}$, energy resolution $\sigma_E=0.063(2)$~meV and $\chi^2=200.7$. \\
Fixing the energy resolution to the calculated resolution; $\sigma_{E \, proj} = 0.0786(8)$~meV and $\sigma_{E \, at \, gap} = 0.0369(4)$~meV, we fit the gaps once again. For the projected energy resolution, we get an overestimated gap size; $1.097(2)$~meV and a slightly larger $\chi^2=277.2$. For the intrinsic energy resolution at the gap, we get a smaller gap size, $1.066(2)$~meV, with a much larger $\chi^2=395.2$. The energy resolution found by the fitting function lies between the calculated projected and intrinsic resolution values at the gap. This shows us that the ''effective`` Q-resolution lies between our two extreme cases and that the choice of energy resolution influences the gap size obtained from the fit.

\end{document}